\documentclass[prl,aps,twocolumn,tightenlines,notitlepage,nofootinbib,preprintnumbers,letterpaper,superscriptaddress]{revtex4-1} 
\pdfoutput=1
\usepackage{epsfig}
\usepackage[normalem]{ulem}
\usepackage{amsfonts}
\usepackage{amsmath,amssymb}
\usepackage{graphicx}
\usepackage{url}
\usepackage{color}
\usepackage{multirow}
\usepackage{placeins}
\usepackage[dvipsnames]{xcolor}
\usepackage{epstopdf}
\usepackage{physics,amssymb,amsmath,slashed,color,graphicx,bm}
\usepackage[compat=1.1.0]{tikz-feynman}
\usepackage[T1]{fontenc}
\usepackage{relsize}
\usepackage{hyperref}
\hypersetup{allcolors=[rgb]{0.0 0.0 0.6},linkcolor=[rgb]{0.8 0.0 0.5}}

\usepackage{comment}

\hypersetup{colorlinks,citecolor= nicegreen,linkcolor= nicered}
\definecolor{nicered}{rgb}{0.7,0.1,0.1}
\definecolor{nicegreen}{rgb}{0.1,0.5,0.1}

\newcommand{\lo}[1]{\ensuremath{_{\textrm{#1}}}} 
\newcommand{\eqn}[1]{
	\begin{align}
		#1
	\end{align}
}
\def\MPIK{Max-Planck-Institut f\"ur Kernphysik, Saupfercheckweg 1, 69117 Heidelberg, Germany}
\def\Aachen{Department of Physics, RWTH Aachen University, Otto-Blumenthal-Straße, 52074 Aachen, Germany}

\begin{document}

\title{Sterile neutrino dark matter production in presence of non-standard neutrino self-interactions: an EFT approach}

\author{Cristina Benso} \email{cristina.benso@mpi-hd.mpg.de}
\affiliation{\MPIK} 
\author{Werner Rodejohann} \email{werner.rodejohann@mpi-hd.mpg.de}
\affiliation{\MPIK} 
\author{Manibrata Sen} \email{manibrata.sen@mpi-hd.mpg.de}
\affiliation{\MPIK} 
\author{Aaroodd Ujjayini Ramachandran} \email{aaroodd.ujjayini.ramachandran@rwth-aachen.de}
\affiliation{\Aachen}

\begin{abstract}
\noindent 
Sterile neutrinos with keV-scale masses are popular candidates for warm dark matter. In the most straightforward case they are produced via oscillations with active neutrinos. We introduce effective self-interactions of active neutrinos and investigate the effect on the parameter space of sterile neutrino mass and mixing. Our focus is on mixing with electron neutrinos, which is subject to constraints from several upcoming or running experiments like  TRISTAN, ECHo, BeEST and HUNTER. Depending on the size of the self-interaction, the parameter space moves closer to, or further away from, the one testable by those future experiments. In particular, we show that phase 3 of the HUNTER experiment would test a larger amount of parameter space in the presence of self-interactions than without them.  
We also investigate the effect of the self-interactions on the free-streaming length of the sterile neutrino dark matter, which is important for structure formation observables.  
\end{abstract}

\preprint{}

\maketitle 

\section{Introduction}
\label{sec:intro}
\noindent 
The nature of dark matter (DM), after many decades of investigation, still lies beyond our knowledge. A number of candidates belonging to the class of weakly interacting massive particles (WIMPs) \cite{Arcadi:2017kky} have received a lot of attention from the theoretical and experimental point of view.
However, so far, none of the attempts to capture incontestable evidence of the existence of WIMP DM has given positive results.
This situation makes other species of DM candidates more and more appealing  \cite{Feng:2010gw}. 
Among those, sterile neutrinos with a mass of the order of a few keV proved to have all the credentials to play the role of dark matter \cite{Abazajian:2001nj,Abazajian:2017tcc,Drewes:2016upu}, in addition to the fact that their existence could be motivated also by reasons of symmetry and aesthetics of the Standard Model (SM), and by the need of giving mass to the SM active neutrinos.

One of the main differences between sterile neutrinos and WIMPs resides in the way in which they have been produced in the early Universe. 
While WIMPs would have been thermally produced in great abundance and reached equilibrium with the primordial plasma before freezing-out due to the cooling and expansion of the Universe, the most natural way for sterile neutrinos to be produced is through neutrino oscillation and collisions in the primordial plasma, as described first by Dodelson and Widrow (DW) \cite{Dodelson:1993je}. 
In this scenario, one or more active neutrino states can mix with a sterile state $\nu_s$, which is our DM candidate. During the propagation, neutrinos scatter with particles in the plasma through weak interactions, and this causes the mixed state to collapse to a weak interaction eigenstate (active neutrino), thereby giving origin to a non-negligible abundance of frozen-in sterile neutrinos.

In the DW mechanism, the parameters that determine the final abundance of the sterile neutrino DM are the mass of the sterile neutrino $m_s$ and its mixing with the active species $\theta$. 
The case in which the entire content of dark matter is constituted by sterile neutrinos is represented by a straight line in this parameter space.
More complicated mechanisms of production, such as the Shi-Fuller mechanism \cite{Shi:1998km} which describes resonant production, or non-standard cosmological or particle physics scenarios in which the production occurs \cite{Gelmini:2004ah,Bezrukov:2009th,Nemevsek:2012cd,Berlin:2016bdv,Bezrukov:2017ike,Hansen:2017rxr,Johns:2019cwc,Benso:2019jog, Gelmini:2019wfp}, depend on more parameters and modify the region of the parameter space where the condition $\Omega_s\,h^2= \Omega_{\text{DM}}\,h^2= 0.12$ \cite{Planck:2018vyg} is satisfied. 
In the standard DW scenario, the values of $m_s$ and $\sin^2(2\theta)$ that correspond to $\Omega_s\,h^2= \Omega_{\text{DM}}\,h^2= 0.12$, are ruled out by  constraints coming from X-ray observations \cite{Perez:2016tcq, Ng:2019gch}. 
 However, non-standard self-interactions (NSSI)  can assist in efficient production of sterile neutrinos in such a way that the parameter space region originating the correct abundance of sterile neutrinos is enlarged and the X-ray constraints evaded, which has been demonstrated for Dirac neutrinos in Refs.\ \cite{DeGouvea:2019wpf, Kelly:2020pcy,Kelly:2020aks}. 

The purpose of our work is to study these NSSI induced sterile neutrino dark matter production in a model-independent framework. Using the formalism of effective field theory (EFT) to incorporate neutrino NSSI, we study the production of sterile neutrino dark matter through the Dodelson-Widrow mechanism. We study here generic flavor-diagonal NSSI for Majorana neutrinos; they can be mediated by a scalar,  pseudoscalar or an axial-vector particle.  Focusing only on one type of NSSI at a time, we discover that the interplay between the NSSI contribution to the interaction rate and the one to the thermal potential leads to a widening of the allowed parameter space in the direction of smaller, as well as larger mixing angles. We have not limited our study to cases where the sterile neutrino constitutes the entire DM relic density.
We also show that in the case of such a ``cocktail DM'', where sterile neutrinos forms only a certain percentage of the entire DM abundance, the allowed parameter space in the presence of NSSI is even wider owing to a relaxation of the X-ray constraints. Finally, we point out that the extra NSSI among the neutrinos needed for efficient production of DM does not cause problems with respect to the current limits from structure formation. Our focus is on mixing of the sterile neutrinos with electron neutrinos. 
Upcoming experiments such as TRISTAN \cite{KATRIN:2018oow} (an update of KATRIN), ECHo \cite{Gastaldo:2017edk}, HUNTER \cite{Martoff:2021vxp} and BeEST \cite{Fretwell:2020ntq} will  be searching for  keV-scale sterile neutrinos using beta decays or electron capture, and therefore any new-physics related  modification of the parameter space with respect to the DW one is of interest. 
In particular, the HUNTER experiment, in its planned third phase, will be able to test a sizable fraction of the parameter space relevant for this work. This makes it an exciting time to study the interplay of astrophysical and terrestrial probes of such NSSI-mediated sterile neutrino DM.

The paper is organized as follows. 
In the \hyperref[sec:DW]{second section}, we introduce the standard Dodelson-Widrow mechanism highlighting the different factors that determine the final abundance of sterile neutrinos, in order to clearly identify where the action of NSSI impacts and in which way. 
In the \hyperref[sec:NSIs]{third section}, we briefly describe the EFT formalism in which we work and present the types of NSSI that we include in the study. 
In the \hyperref[sec:results]{fourth section}, we show the results of our work, comment on where they place in the landscape of near future experiments and with respect to the current constraints coming from different observables. 
Furthermore, we give an outlook on the possible future developments of this work and we \hyperref[sec:conclusion]{conclude} with the summary of what has been covered in our study. 
In the \hyperref[appendix:potential]{appendix} one  can find the explicit calculation of the NSSI contributions to the thermal potentials, carried out in the EFT framework.

\section{The Dodelson-Widrow mechanism}
\label{sec:DW}
\noindent
We introduce in this work a sterile neutrino as an additional keV-scale SM singlet fermion. 
Note that we do not require these keV sterile neutrinos to explain the active neutrino masses. This can, in principle, arise from other heavy sterile neutrinos which are not relevant to this discussion. 
We study the case in which sterile neutrino dark matter $(\nu_s)$ is produced in the early Universe through oscillation and collisions as described in the Dodelson-Widrow mechanism. 
The requirement for this mechanism to work is that the sterile neutrino mixes at least with one active neutrino flavor and, for simplicity, we consider that sterile neutrinos mix only with one active species. Hence, we define an additional mass eigenstate as a linear combination of a sterile and an active state, i.e., $\nu_4 = \cos\theta\, \nu_s + \sin\theta \, \nu_\alpha$. In our case $\alpha = e$, which is motivated mainly by the advent of experiments such as TRISTAN \cite{KATRIN:2018oow} (update of KATRIN), ECHo \cite{Gastaldo:2017edk}, HUNTER \cite{Martoff:2021vxp} and BeEST \cite{Fretwell:2020ntq}, that will be sensitive to signals of such DM candidates if they mix with electron neutrinos or antineutrinos. 
Moreover, although there are no sensitivity studies available yet, experiments such as Project 8 \cite{Guigue:2017wzr} and PTOLEMY \cite{PTOLEMY:2019hkd} will eventually also be sensitive to keV-scale sterile neutrinos. 

The $\nu_s$ population in the early Universe is assumed to be negligible, whereas the active neutrinos are in a thermal bath. 
During the evolution of the Universe, the $\nu_\alpha$ oscillate into $\nu_s$, and at the same time, undergo weak interactions, which reset the active neutrino flavor again.
The sterile neutrinos, on the other hand, are SM singlets, and do not experience any weak forces. 
This interplay between oscillations and collisions allows a non-trivial population of sterile neutrinos to build up.

Mathematically, the time evolution of the $\nu_s$ distribution function $f_{s}(p, t)$ can be calculated from the following semi-classical Boltzmann equation~\cite{Abazajian:2001nj}, 
\begin{widetext}
\begin{align}
    & \frac{\partial}{\partial t} f_{s}(p, t)-  H p \frac{\partial}{\partial p} f_{s}(p, t) \approx  \frac{\Gamma_{\alpha}(p)}{4} \frac{ \Delta^{2}(p) \sin ^{2} (2 \theta)}{\Delta^{2}(p) \sin ^{2} (2 \theta)+ D^{2}(p)+\left[\Delta(p) \cos( 2 \theta)-\mathcal{V}_{\alpha}(p)\right]^{2}}\left[f_{\alpha}(p,t)-f_{s}(p, t)\right].
    \label{eq:Boltzmann}
\end{align}
With some manipulation we can rewrite it in the following shape 
\begin{align}
     \frac{\partial}{\partial T} f_{s}(p, T) & + \frac{1}{T} \left[ 1 + \frac{T}{3} \frac{g'_S}{g_S} \right] p \frac{\partial}{\partial p} f_{s}(p, T) = h(p, T) \left[f_{\alpha}(p,T)-f_{s}(p, T)\right] \nonumber\\
     & \approx \frac{d t}{d T} \frac{\Gamma_{\alpha}(p)}{4} \frac{ \Delta^{2}(p) \sin ^{2} (2 \theta)}{\Delta^{2}(p) \sin ^{2} (2 \theta)+ D^{2}(p)+\left[\Delta(p) \cos (2 \theta)-\mathcal{V}_{\alpha}(p, T)\right]^{2}}\left[f_{\alpha}(p,T) - f_{s}(p, T)\right],
    \label{eq:Boltzmann_T}
\end{align}
\end{widetext}
suitable to obtain the semi-analytical solution that we discuss later in the \hyperref[sec:struc]{subsection} devoted to the impact of NSSI on structure formation. 
Here $\Gamma_{\alpha}$ denotes the net active neutrino interaction rate, $\mathcal{V}_{\alpha}$ measures the effective potential experienced by the neutrino, the quantum damping rate for neutrinos is $D(p)=\Gamma_{\alpha}/2$, and $\Delta \simeq m_s^2/(2p)$ is the vacuum oscillation frequency, dominated by the sterile mass. The function  $h$ contains the details of the Dodelson-Widrow mechanism, whereas $g_S$ denotes the effective degrees of freedom, and $g'_S$ denotes its derivative with respect to temperature.
The initial $\nu_\alpha$ distribution function, $f_{\alpha}$, is taken to be Fermi-Dirac, while the initial $\nu_s$ population can be taken to be negligible. The second term on the LHS describes the expansion of the Universe through the Hubble parameter, $H$. 

Within the SM, the contribution to $\mathcal{V}_{\alpha}$ arises from forward scattering of the neutrinos with the leptons in the plasma. Assuming the Universe is lepton-symmetric and ignoring the tiny baryon asymmetry, finite density contributions to the  effective potential become zero and the resulting thermal potential is given by \cite{Abazajian:2001nj, Abazajian:2005gj} 
\begin{eqnarray}\label{Eq:VT_SM}
\mathcal{V}_{T,\, {\rm SM}} = &-& \frac{8\sqrt{2}\, G_{\rm F} p}{3 M_{\rm Z}^2}
\left(\langle E_{\nu_\alpha} \rangle n_{\nu_\alpha} + \langle
E_{\bar\nu_\alpha} \rangle n_{\bar\nu_\alpha}\right) \cr &-&
\frac{8\sqrt{2}\, G_{\rm F} p}{3 M_{\rm W}^2} \left(\langle E_\alpha
\rangle n_\alpha + \langle E_{\bar\alpha} \rangle
n_{\bar\alpha}\right) \cr &\simeq& - 3.72\,G_F p\,T^4 \left(\frac{2}{M_W^2}+ \frac{1}{M_Z^2} \right),
\end{eqnarray}
where $G_F$ is the Fermi constant, and $M_W,\,M_Z$ are the masses of the $W$ and $Z$ gauge bosons; the number densities of neutrinos and their corresponding charged leptons are denoted by $n_{\nu_\alpha}$ and $n_\alpha$, respectively; their averaged energies are  $\langle E_{\nu_\alpha} \rangle$ and $\langle E_\alpha
\rangle$. 
The thermally averaged scattering rate for $\nu_\alpha$, due to neutrinos scattering off charged leptons, is given by \cite{Abazajian:2001nj, Abazajian:2005gj}
 \begin{equation}\label{Eq:SM_rate}
 \Gamma_{\alpha} =
C_\alpha(T)\, G_F^2 p\,T^4\,.
 \end{equation}  
where $p$ is the sterile neutrino momentum, and  $C_\alpha(T)$ are temperature-dependent functions defined in \cite{Abazajian:2001nj,Asaka:2006nq}.

Plugging these expressions into Eq.\ (\ref{eq:Boltzmann}), one can compute the sterile neutrino relic density in the early Universe.
This is a very elegant mechanism, and can account for the entire observed DM relic density. 
However, this mechanism is in strong tensions with astrophysical searches. 
The non-zero mixing between $\nu_s$ and $\nu_\alpha$ allows the $\nu_s$ to decay radiatively into $\nu_\alpha$ and a photon. 
For $\nu_s$ heavier than a few keV, this can lead to the production of X-rays. 
Non-observation of X-rays by telescopes potentially rules out the viable mixing angles required for the Dodelson-Widrow mechanism to explain the observed DM density. 
On the other hand, lighter sterile neutrinos run into difficulty from phase space considerations in dwarf galaxies.
As a result, the vanilla Dodelson-Widrow mechanism of producing dark matter in the early Universe is currently ruled out. 

More recently, it was suggested that  new, hidden interactions among the active neutrinos can play a major role in alleviating these tensions~\cite{DeGouvea:2019wpf}. This was achieved by postulating secret neutrino interactions mediated by a lepton-number charged scalar~\cite{DeGouvea:2019wpf}, and subsequently, this was also generalised to vector mediators arising out of $U(1)$ gauge extensions of the SM~\cite{Kelly:2020pcy}. These new interactions can be much stronger than the weak interactions, and hence help in effectively producing sterile neutrino dark matter in the early Universe.

In this work, we take this  idea further, and consider neutrino self-interactions in an   effective field theory framework, considering all possible types of NSSI among the active Majorana neutrinos. In the next section, we outline the details of the EFT framework used in this analysis, and the extra contributions to the thermal potential, and the scattering rates that come about because of these new interactions.

\section{NSSI impact on sterile neutrino production}
\label{sec:NSIs}
\noindent
In this section, we outline the EFT framework used to incorporate NSSI into the SM. 
We consider generalised scalar, pseudoscalar and axial-vector neutrino self-interactions.  
In order to remain agnostic about the nature of the high-scale physics, we employ an EFT framework to study the impact of these new interactions.  

Nevertheless, we need to take the underlying mediator of the NSSI into account, if we want to capture the physically correct temperature dependence of the thermal potential. We illustrate this with a Yukawa-like interaction 
\eqn{
\mathcal{L}\lo{int} \supset\lambda_{\phi} \bar{\nu} \mathcal{O} \nu \phi +\mathrm{h.c.} \,
}
between active neutrinos $\nu$ and a mediator  $\phi$. The most generalized NSSI Lagrangian for an active neutrino up to first order in $\Box$ can be derived as (see the  \hyperref[appendix:potential]{appendix})
\begin{eqnarray}
\label{eqn:generallagrangian}
\mathcal{L}_j &=& \frac{G_\phi }{\sqrt{2}} 
\left[ 
(\bar{\nu} \mathcal{O}_j \nu )(\bar{\nu} \bar{\mathcal{O}_j} \nu) - (\bar{\nu} \mathcal{O}_j \nu )\frac{\Box}{m_\phi^2}(\bar{\nu} \bar{\mathcal{O}_j} \nu)\right] \\
&=&
\frac{G_F \epsilon_{j} }{\sqrt{2}} 
\left[ 
(\bar{\nu} \mathcal{O}_j \nu )(\bar{\nu} \bar{\mathcal{O}_j} \nu) - (\bar{\nu} \mathcal{O}_j \nu )\frac{\Box}{m_\phi^2}(\bar{\nu} \bar{\mathcal{O}_j} \nu)\right]
,
\end{eqnarray}
where $G_{\phi}=\frac{\sqrt{2}\lambda_{\phi}^2}{m_\phi^2}$ is the non-standard interaction strength in terms of coupling constant $\lambda_\phi$ and mediator mass $m_\phi$. We have further defined 
$G_{\phi}= G_F \epsilon$, where $\epsilon$ is used to indicate the NSSI strength compared to the standard weak interactions. 
Here $\mathcal{O}_j,\bar{\mathcal{O}_{j}}$ are elements of a complete set of bilinear covariants, which in general is $\{\mathbb{I}, \gamma^\mu, i\gamma^5, \gamma^\mu\gamma^5, \sigma^{\mu\nu}\}$. For the context of this paper we deal with Majorana neutrinos and flavor-diagonal NSSI, hence $\mathcal{O}=\gamma^\mu, \sigma^{\mu\nu}$ do not contribute, and furthermore we have $\epsilon>0$. 
The second term in the Lagrangian Eq.\ \eqref{eqn:generallagrangian} gives momentum-dependent Feynman rules and is essential to capture the physically correct temperature dependence of the thermal potential, see below and the \hyperref[appendix:potential]{appendix} for details.

The DW mechanism predicts that maximal production of DM happens around temperatures of $T\simeq 133 \, (m_s/{\rm keV})^\frac 13 \,{\rm MeV}$. 
Hence, for the EFT to be valid, it makes sense to have heavy mediators, with masses greater than a few GeV. Furthermore, as long as the mediators are lighter than the $W,Z$ bosons, the strength of the NSSI can be larger than the weak interactions. 
Note that direct bounds on effective NSSI operators are very loose, the strongest bounds come from  contributions to the invisible $Z$-decay~\cite{Bilenky:1999dn}.

\begin{figure}[!t]
\begin{minipage}[c]{.4\textwidth}
\resizebox{\textwidth}{!}{%
\begin{tikzpicture}[baseline=(m)]
    \begin{feynman}
      \vertex[dot] (m) {};
      \vertex [left =of m] (a) {$\nu$};
      \vertex [right =of m] (b)  {$\nu$};
      \node [dot,above =0.7cm of m] (c) ;
     \coordinate [above =1cm of c] (d) ;
      \diagram* {
        (a) -- [fermion,edge label=$k$] (m) -- [fermion,edge label=$k$] (b),
        (m) -- [dashed] (c)[large,dot] , (c) -- [half right] (d) -- [half right, fermion,edge label'=$p$](c),
      };
    \end{feynman}
  \end{tikzpicture} 
 \begin{tikzpicture}[baseline=(m)]
    \begin{feynman}
      \node[dot] (b) ;
      \vertex [left =of b] (m) {$\nu$};
      \node [dot,right =of b] (c) ;
      \vertex [right =of c] (d)  {$\nu$};
      \diagram* {
        (m) -- [fermion,edge label=$k$] (b) --[dashed,edge label'=$k-p$]  (c)-- [fermion,edge label=$k$] (d) ,
        (b) -- [fermion, half left, looseness=2,edge label=$p$]  (c),
      };
    \end{feynman}
  \end{tikzpicture}  
}%
\end{minipage}
	\caption{Self-energy diagrams relevant for the thermal potential, tadpole (left) and sunset (right).}
	\label{fig:PotentialDiagrams}
\end{figure}
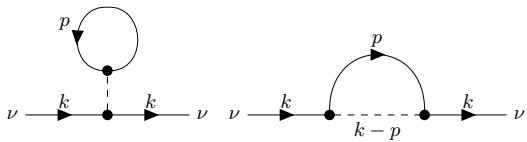 

The NSSI would lead to additional channels for producing sterile neutrinos in the early Universe. This gives an extra contribution to Eq.\ (\ref{eq:Boltzmann}) through the interaction rate $\Gamma$ and the forward-scattering potential $\mathcal{V}_\alpha$. 
The computation of the thermal scattering rate is straightforward and we performed it following \cite{Kelly:2020pcy, Paraskevas:2018mks, Pal:2010ih} and using the FeynCalc tool \cite{MERTIG1991345,Shtabovenko:2016sxi,Shtabovenko:2020gxv} to calculate the amplitudes of the processes.
There is, as mentioned above, a slight subtlety involved while computing  the potential. The contribution to the potential arises from the self-energy diagrams depicted in FIG.\ \ref{fig:PotentialDiagrams}. 
The tadpole diagram gives a contribution proportional to the lepton asymmetry, and can be conveniently ignored if we assume a fairly lepton symmetric Universe. 
Now, if one integrates out the mediator in the sunset diagram (right panel of FIG.\ \ref{fig:PotentialDiagrams}), there is no way to introduce a momentum dependence in the self-energy, since the dependence comes from the momentum of the mediator.
As a result, the EFT need to be expanded to higher orders to compute the potential, see Eq.\ (\ref{eqn:generallagrangian}). 
Taking these terms into account, the contribution to the
thermal potential is given by 
\begin{equation}
\mathcal{V}_{\rm NSSI} = - \frac{7\sqrt{2}\pi^2 }{45m_\phi^2} G_{F} p T^4 \times
\left\{ 
\begin{array}{cl} 
\epsilon_{S,P}\,\,,& \hspace{0.5cm} \mathcal{O}=\mathbb{I},\,i\gamma^5 \,,\\
2 \epsilon_A\,\, ,&\hspace{0.5cm} \mathcal{O}=\gamma^\mu \gamma^5 \,.\\
\end{array}\right.
\label{eq:NSSIPotential}
\end{equation}
For the interaction rate, for which the zeroth order EFT is sufficient, the result is 
\begin{equation}
\Gamma_{\rm NSSI} = \frac{7\pi}{180 }  G_F^2 p T^4  \times
\left\{ 
\begin{array}{cl} 
\epsilon_{S,P}^2\,\,,& \hspace{0.5cm} \mathcal{O}=\mathbb{I},\,i\gamma^5 \,,\\
\frac{4}{3}\epsilon_A^2\,\,,&\hspace{0.5cm} \mathcal{O}=\gamma^\mu\gamma^5 \,.\\
\end{array}\right.
\label{eq:NSSIRates}
\end{equation}
These additional contributions  and help in efficient production of sterile neutrinos in the early Universe.

\section{Results and discussion}
\label{sec:results}
\noindent
In this section, we discuss the results for the production of sterile neutrino dark matter in the presence of NSSI\footnote{The code we wrote and used to get these results is publicly available at \url{https://github.com/cristinabenso92/Sterile-neutrino-production-via-Dodelson-Widrow-with-NSSI}.}.

\subsection{Sterile neutrino parameter space}

\begin{figure*}[hp!]
  \begin{tabular}{cc}
    \includegraphics[width=.46\textwidth]{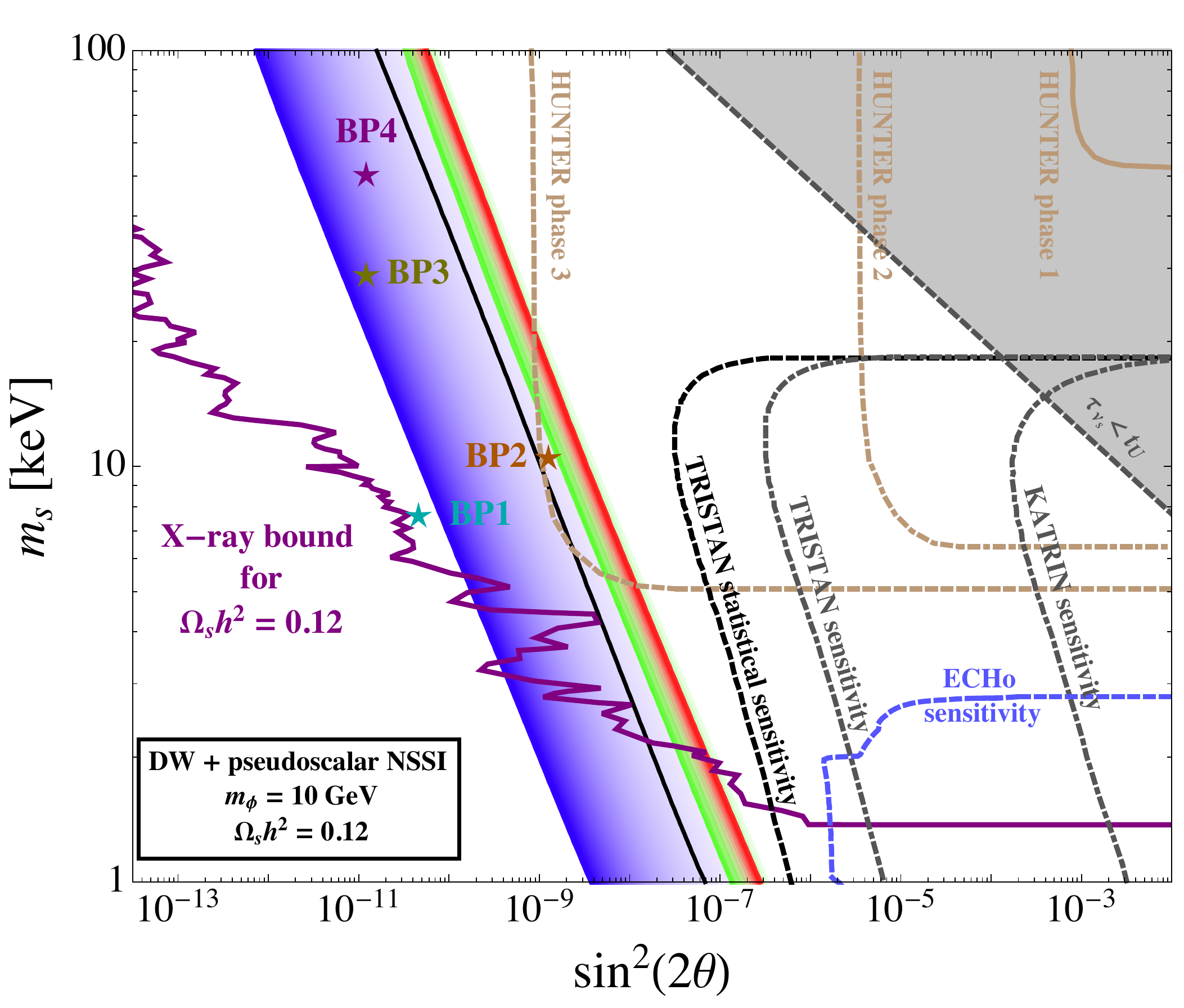} &
    \includegraphics[width=.51\textwidth]{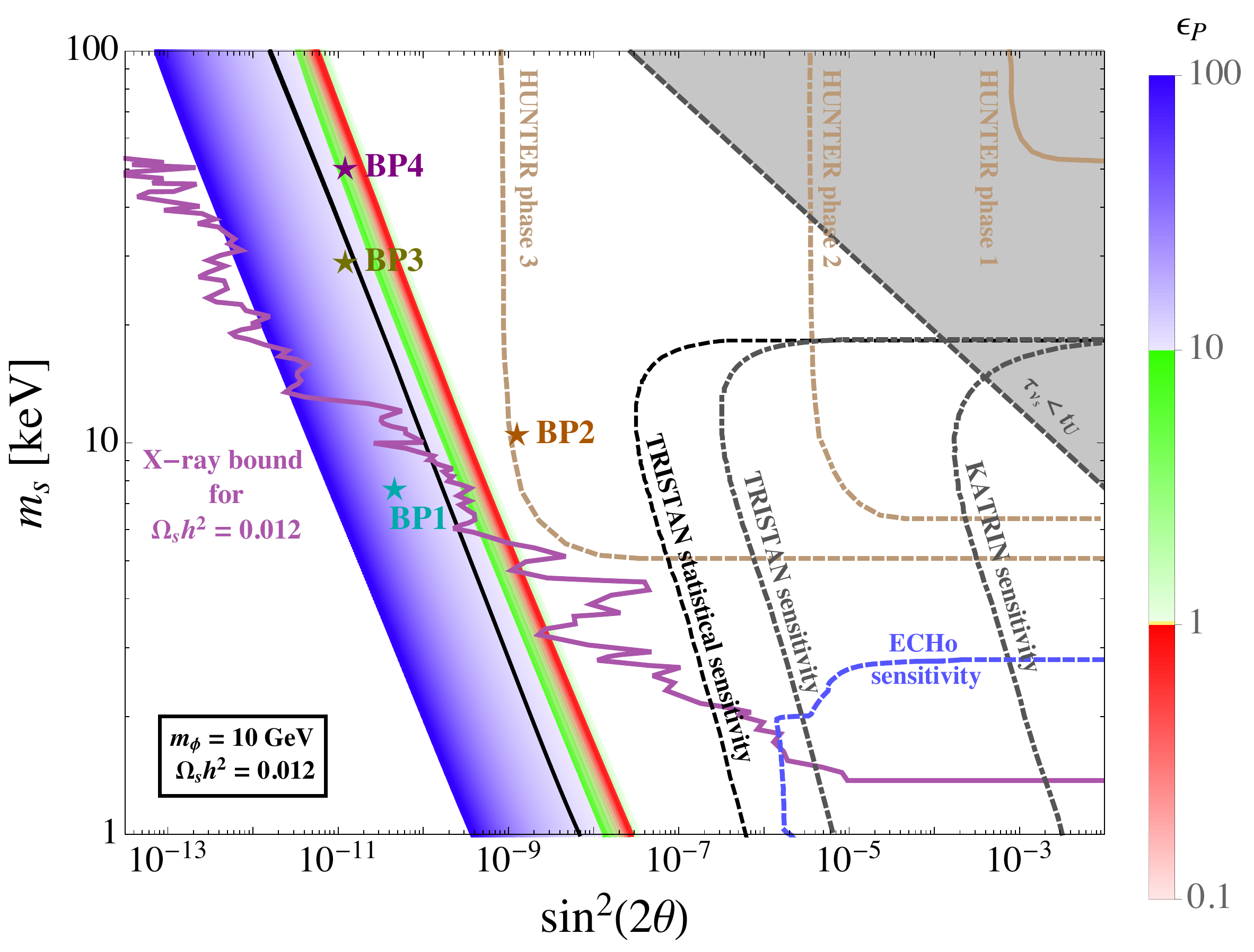}\\
    \includegraphics[width=.46\textwidth]{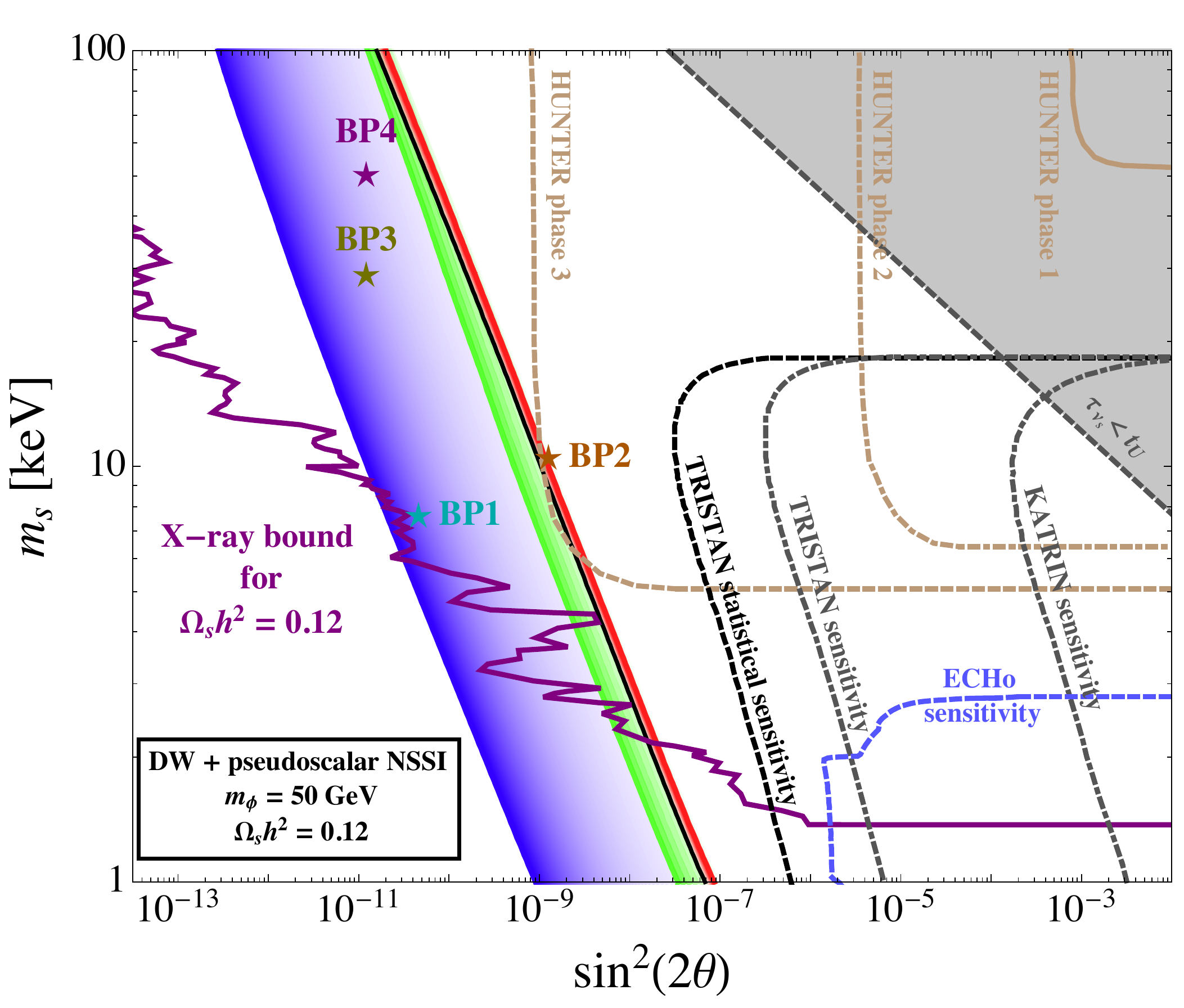}  &
    \includegraphics[width=.51\textwidth]{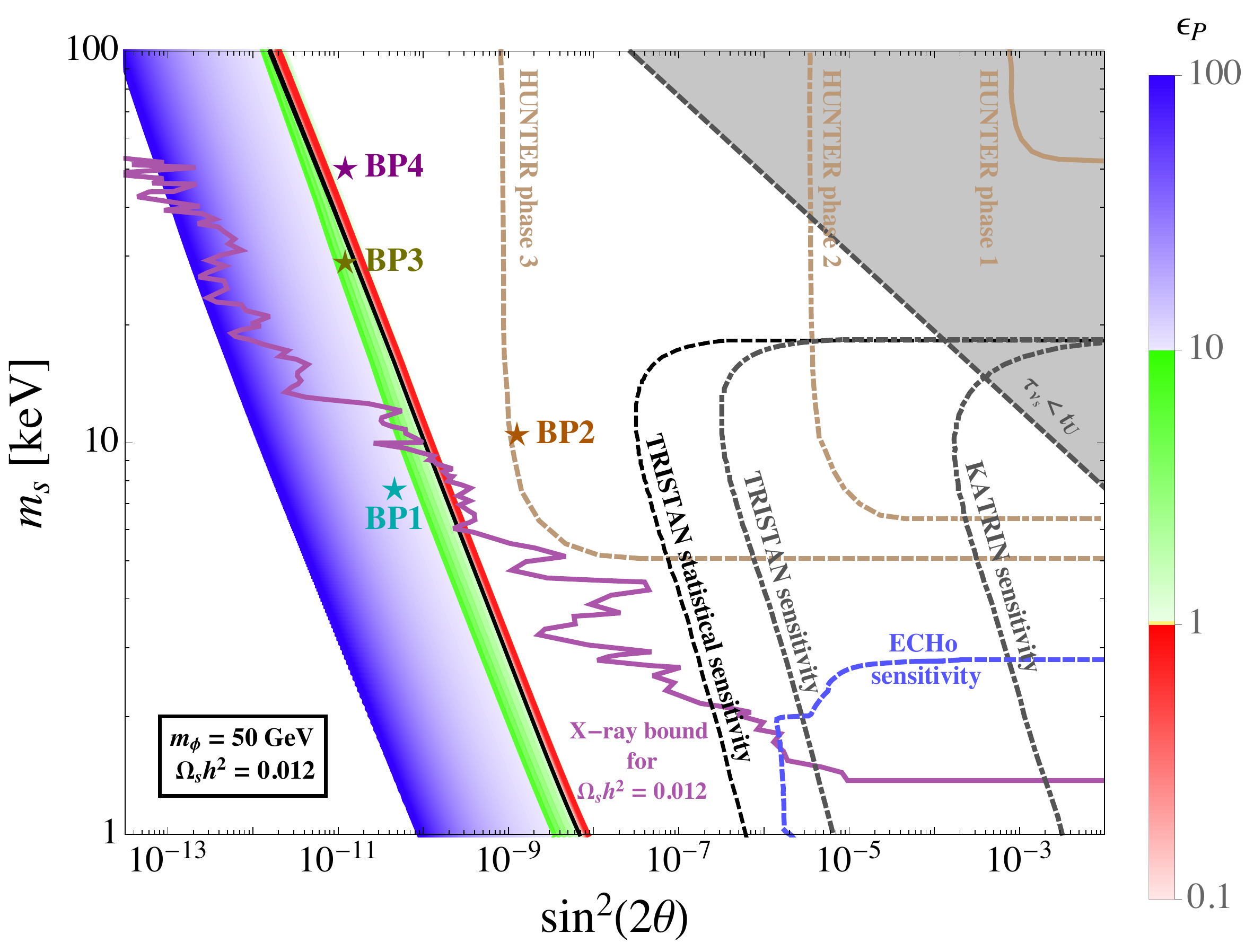} \\
    \includegraphics[width=.46\textwidth]{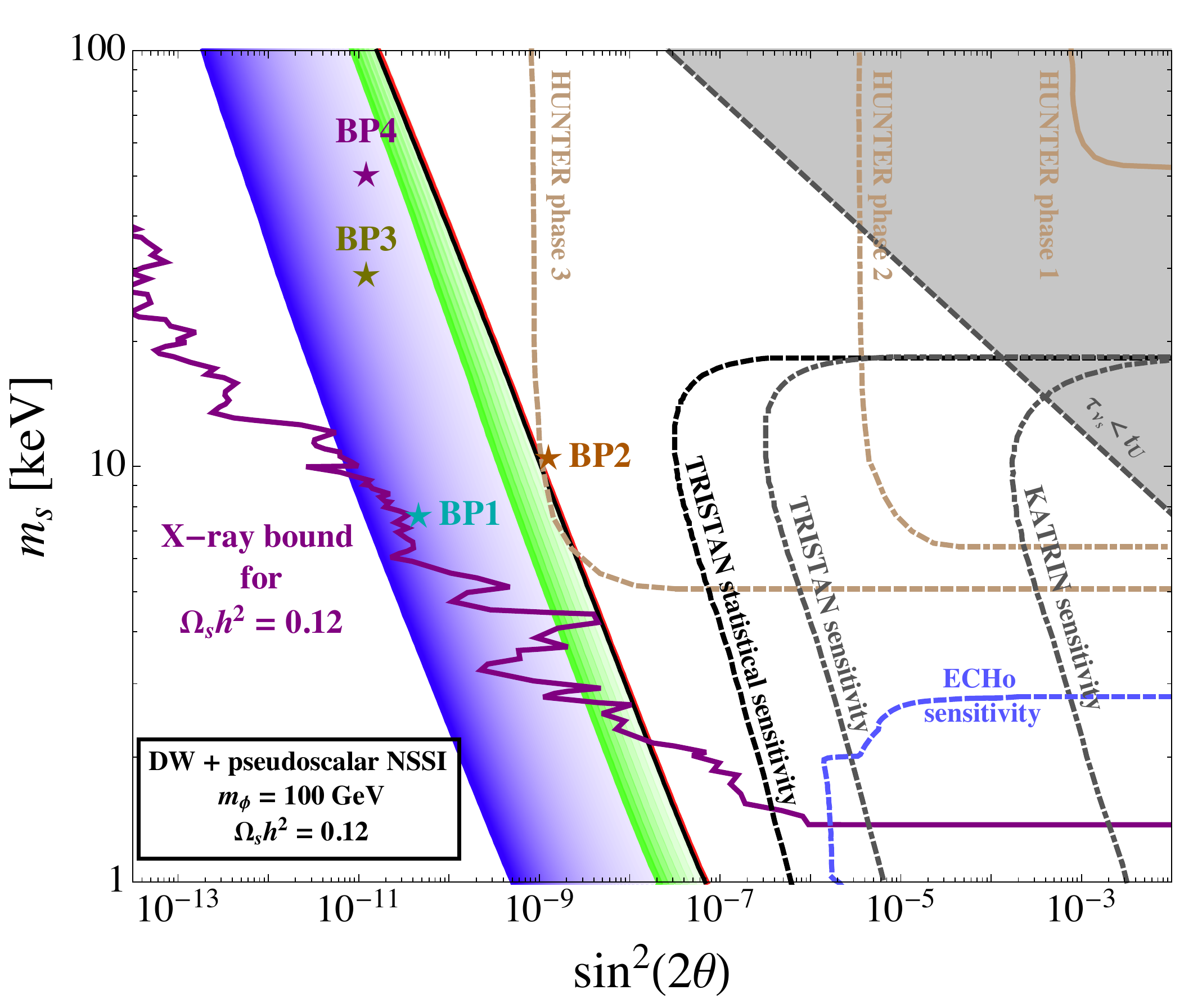} &
    \includegraphics[width=.51\textwidth]{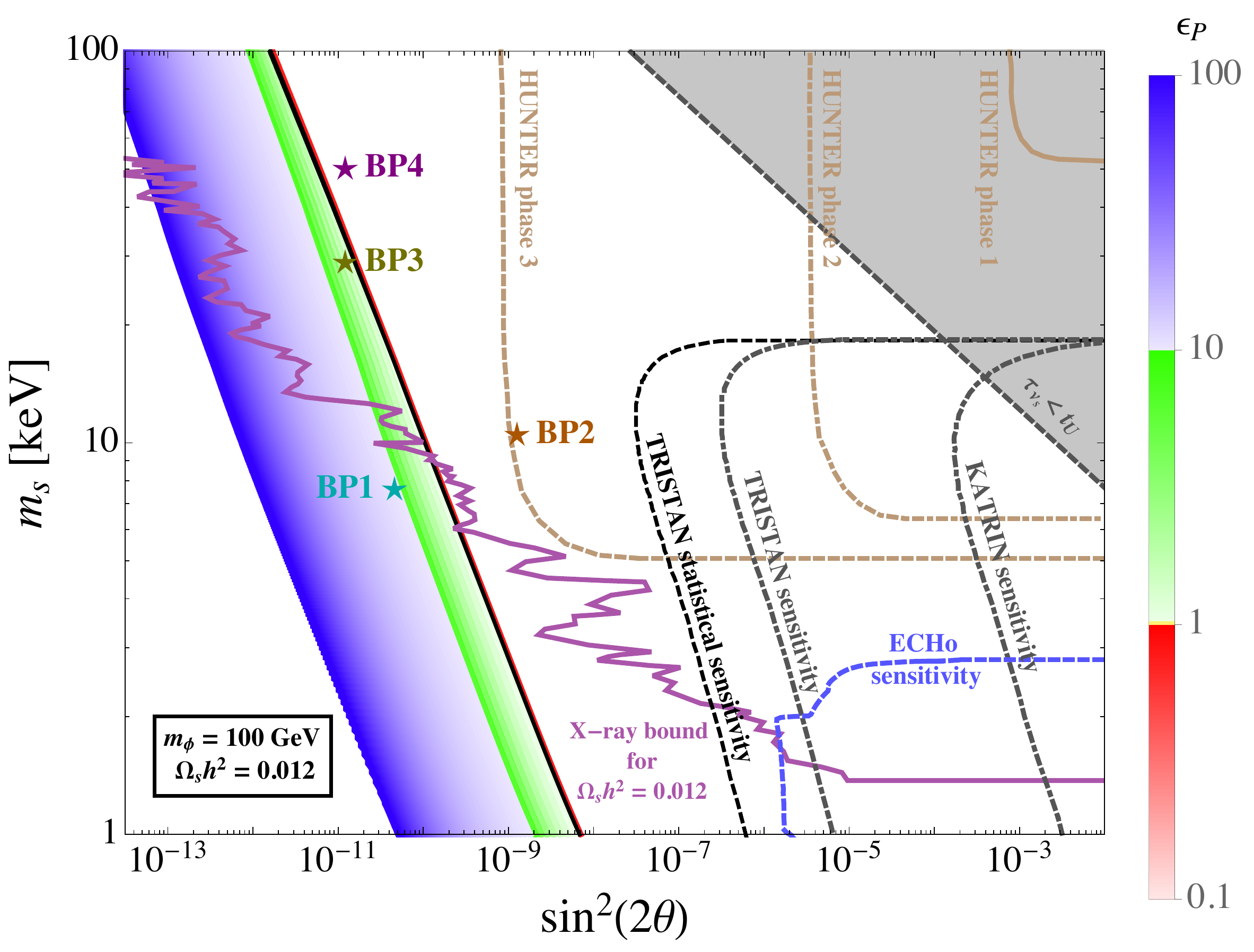} \\ 
  \end{tabular}
  \caption{Pseudoscalar mediator case. 
   The oblique straight lines are constituted by the points corresponding to the values of $m_s$ and $\sin^2(2\theta)$ for which the condition $\Omega_{s} h^2 = \Omega_{\text{DM}} h^2 = 0.12$ (left panel) and 
   $\Omega_{s} h^2 = 0.1 \times \Omega_{\text{DM}} h^2 = 0.012$  (right panel) 
  is satisfied, for different strengths of NSSI parameterized by $\epsilon_P$. 
   The black line corresponds to the case of standard Dodelson-Widrow production, i.e.\ $\epsilon_P = 0$, while different shades of red, green and blue correspond to increasing values of $\epsilon_P$ in the range [0.1, 100]. 
   The regions in which upcoming experiments will be sensitive  are enclosed by beige (HUNTER), black and gray (TRISTAN) and blue (ECHo) lines.
  The purple line represents the current constraint from X-ray observations. Three values of mediator mass were chosen, $m_\phi = 10, 50, 100$ GeV for the upper, middle and lower panel. 
  Four benchmark points BP1-4 are also given, see main text. The scalar mediator case looks identical. } 
\label{fig:pseudoscalar}
\end{figure*}

\begin{figure*}[hp!]
  \begin{tabular}{cc}
    \includegraphics[width=.46\textwidth]{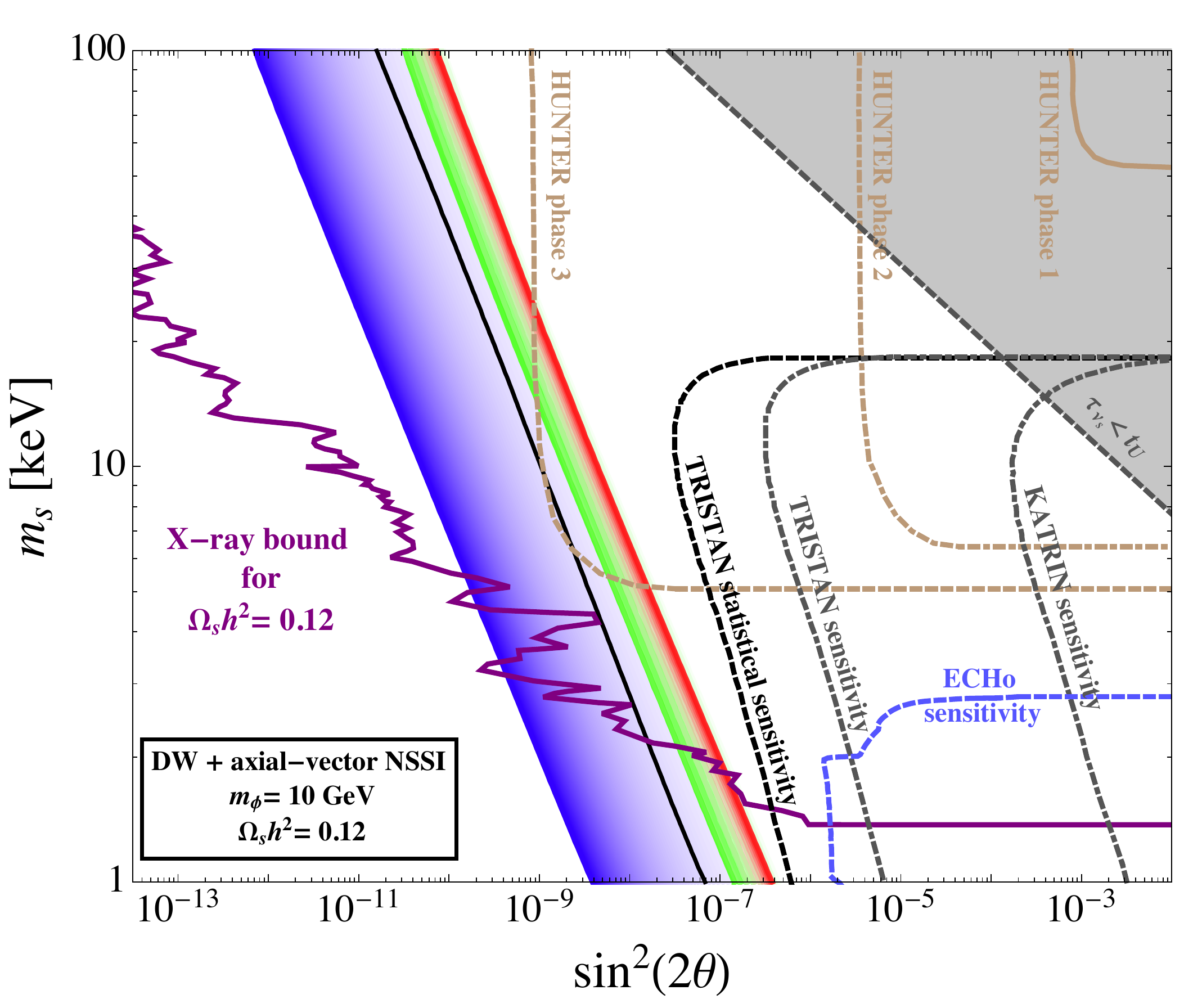} &
    \includegraphics[width=.51\textwidth]{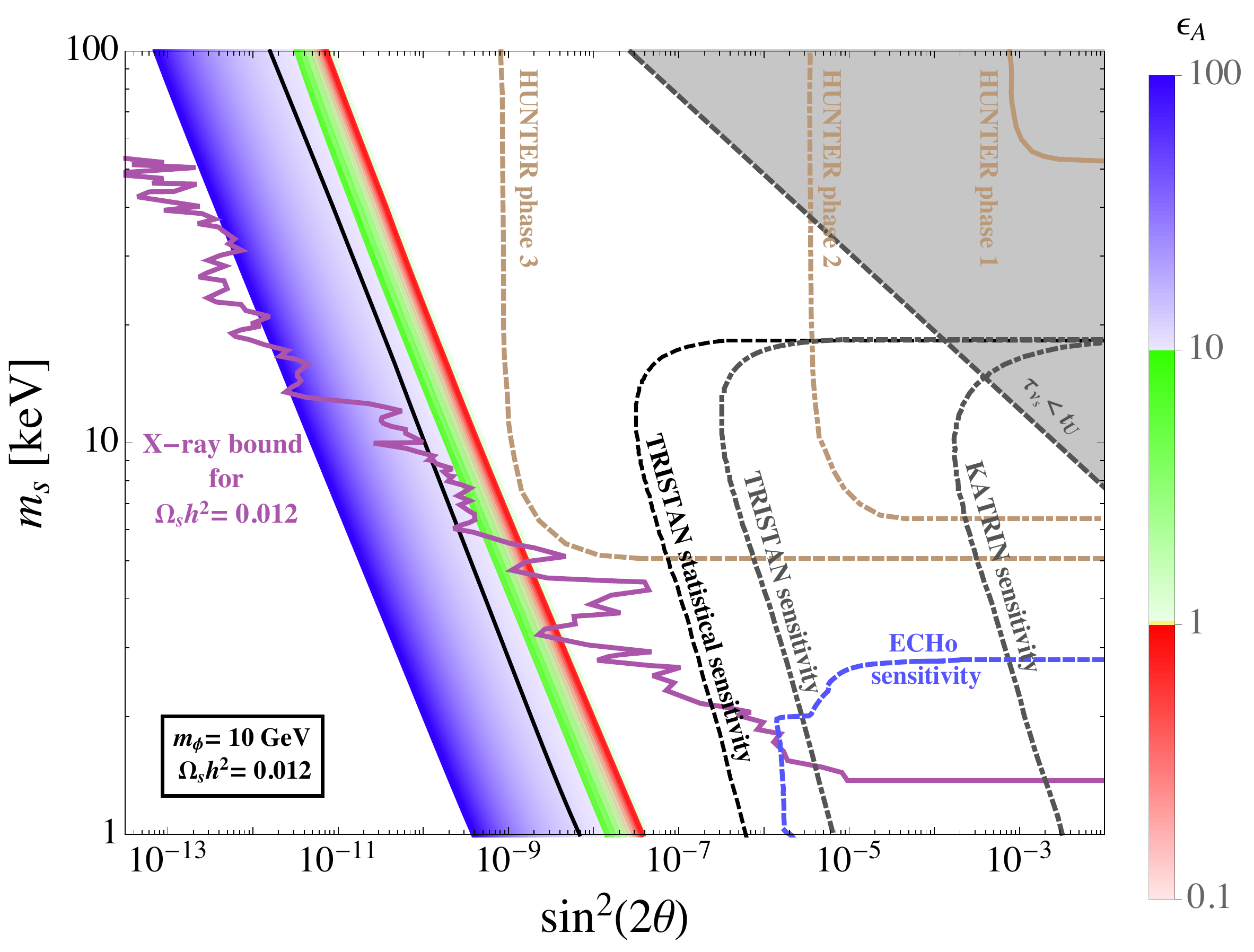}\\
    \includegraphics[width=.46\textwidth]{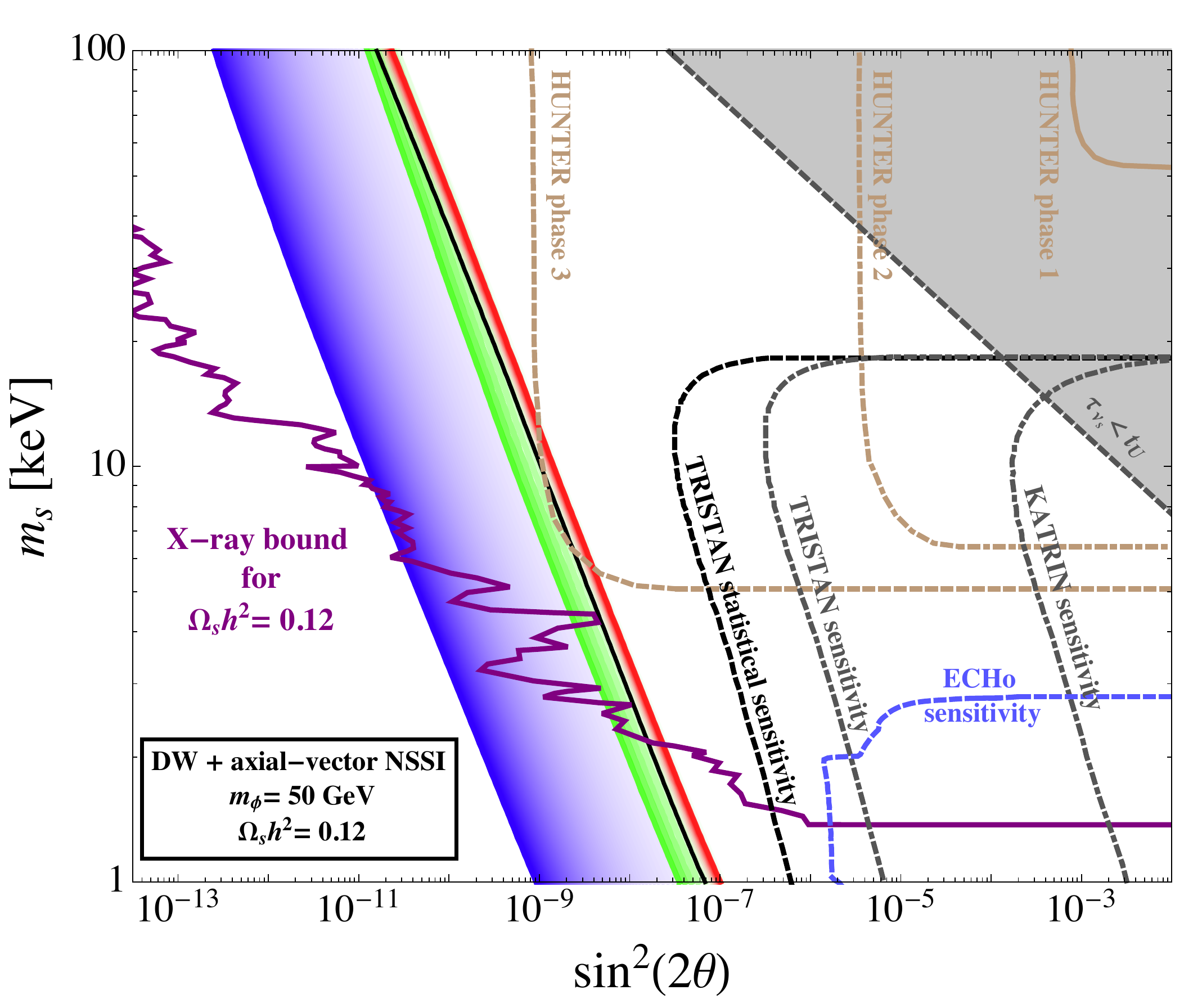}  &
    \includegraphics[width=.51\textwidth]{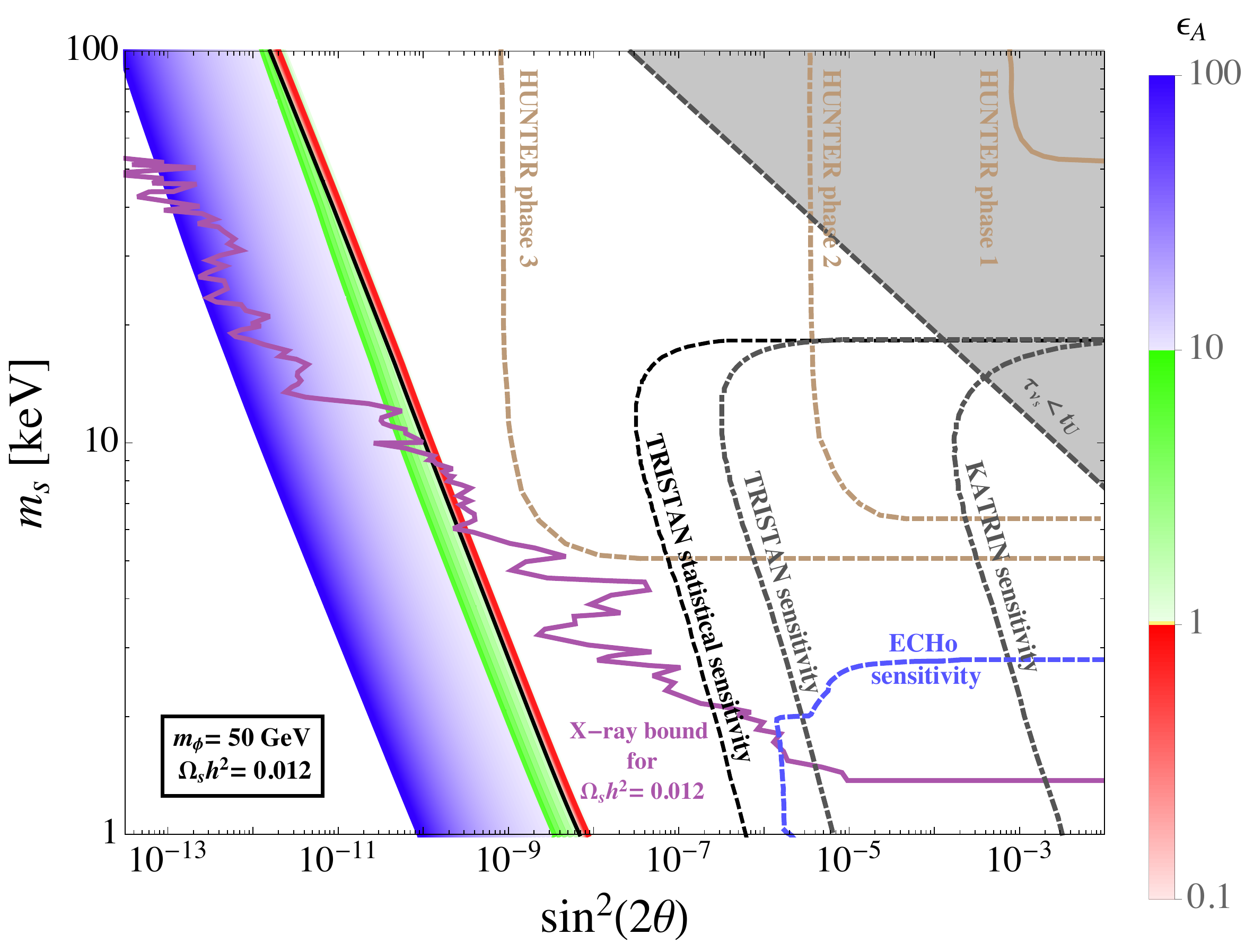} \\
    \includegraphics[width=.46\textwidth]{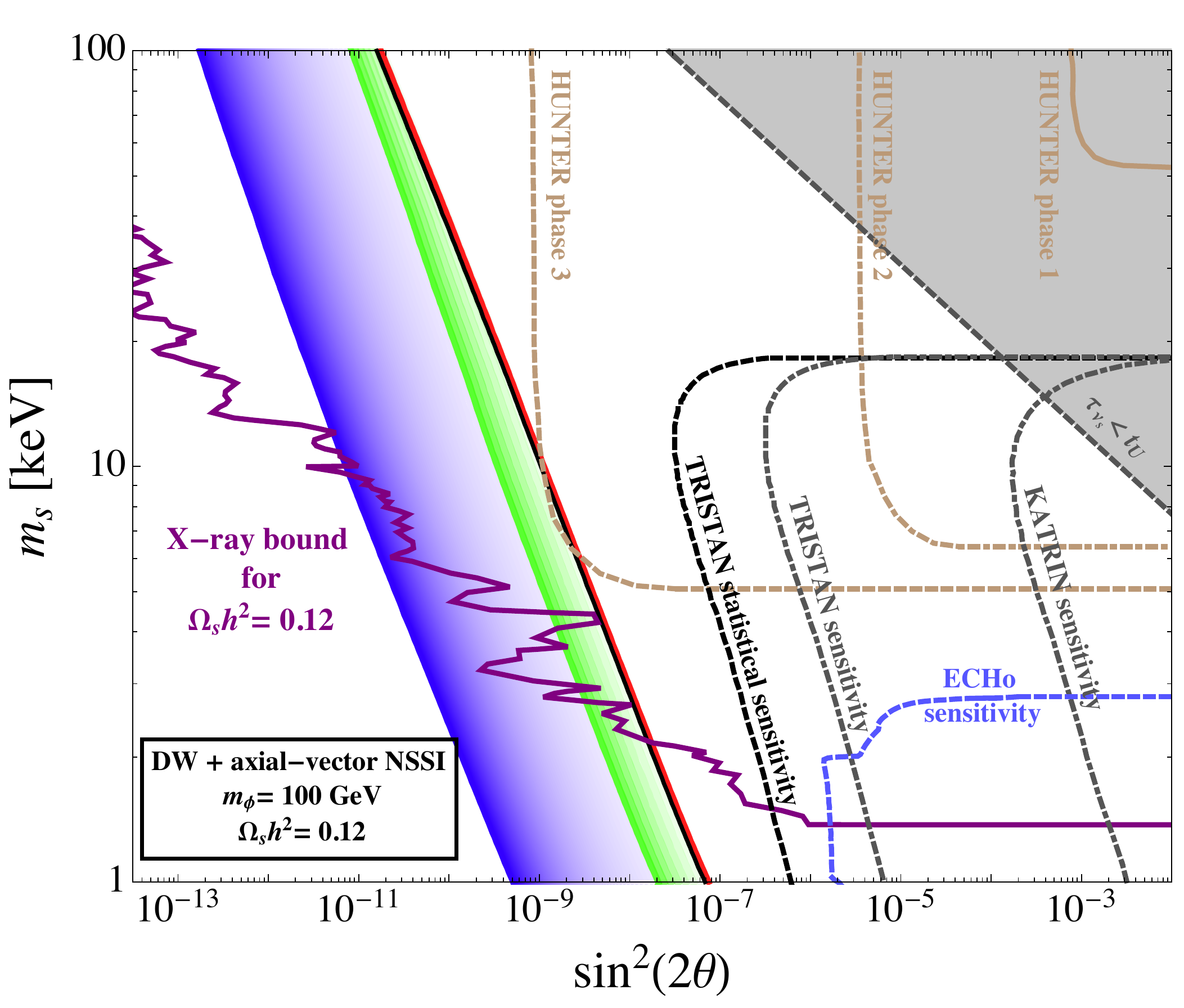} &
    \includegraphics[width=.51\textwidth]{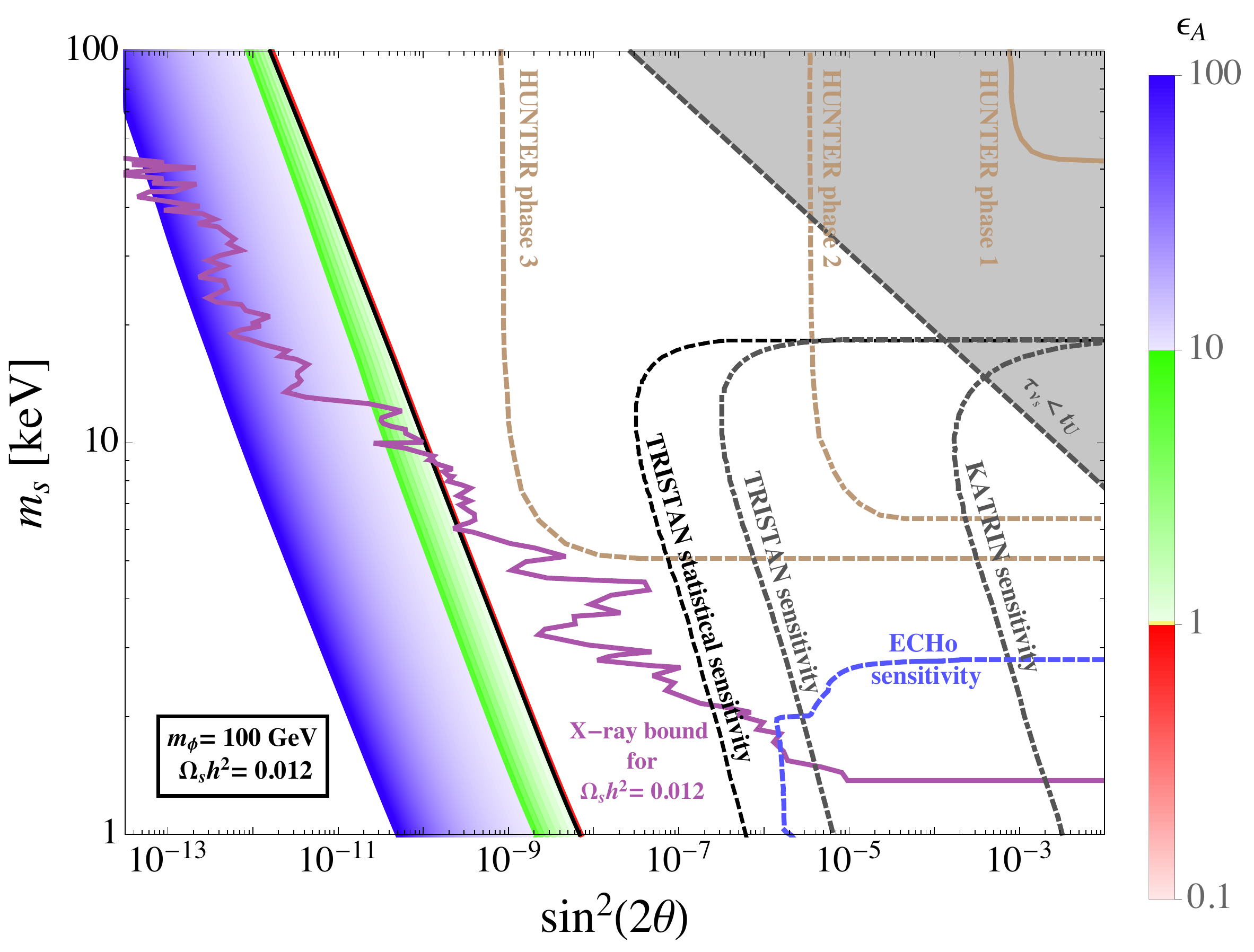} \\ 
  \end{tabular}  
  \caption{\small Axial-vector mediator case. The oblique straight lines are constituted by the points corresponding to the values of $m_s$ and $\sin^2(2\theta)$ for which the condition $\Omega_{s} h^2 = \Omega_{\text{DM}} h^2 = 0.12$ (left panel) and 
   $\Omega_{s} h^2 = 0.1 \times \Omega_{\text{DM}} h^2 = 0.012$  (right panel) 
  is satisfied, for different strengths of NSSI parametrized by $\epsilon_A$. 
   The black line corresponds to the case of standard Dodelson-Widrow production, i.e.\ $\epsilon_A = 0$, while different shades of red, green and blue correspond to increasing values of $\epsilon_A$ in the range [0.1, 100]. 
   The regions in which upcoming experiments will be sensitive  are enclosed by beige (HUNTER), black and gray (TRISTAN) and blue (ECHo) lines.
  The purple line represents the current constraint from X-ray observations. Three values of mediator mass were chosen, $m_\phi = 10, 50, 100$ GeV for the upper, middle and lower panel. }
\label{fig:axial-vector}
\end{figure*}

\noindent 
FIGS.\ \ref{fig:pseudoscalar} and  \ref{fig:axial-vector} summarize the impact of respectively pseudoscalar and axial-vector NSSI on the production of sterile neutrinos in the early Universe, noticeable in their abundance today. 
We do not show the case of scalar NSSI because we found that its effect on the thermal potential and the interaction rate of active neutrinos and, consequently, on the final abundance of sterile neutrinos is indistinguishable from the effect of pseudoscalar NSSI. The differences between these two kinds of interactions emerge only at the level of UV complete models, but the investigation of such models goes beyond the purpose of our work. 
In the left panels  of both figures, we see that the region of the parameter space that satisfies the condition $\Omega_{s} h^2 = \Omega_{\text{DM}} h^2 = 0.12$ is broadened by the presence of NSSI: depending on the strengths $\epsilon_{P, A}$ of the NSSI, we obtain a band instead of the black line representing the standard Dodelson-Widrow case.
Furthermore, we observe that, depending on the mass of the NSSI mediator, the effect of NSSI shifts the DW line towards smaller or larger mixing angles in a different way. 
In particular, for the lightest mediator considered ($m_\phi=10$ GeV), the effect of NSSI with small $\epsilon_{P, A}$ is to suppress the sterile neutrino production leading to larger values of the mixing angle needed to reach $\Omega_{s} h^2 = \Omega_{\text{DM}} h^2$. 
On the other hand, for $\epsilon_{P, A} > 2$, the effect of NSSI goes in the opposite direction enhancing the production of sterile neutrinos and shifting the DW line towards smaller values of the mixing angle. This happens because as $\epsilon$ switches on, there is a new contribution to the interaction rate $\Gamma_\alpha$ and the potential $\mathcal{V}_\alpha$, which is proportional to 
$\epsilon^2$ and $\epsilon$, respectively. 
Initially there is a suppression of $\nu_e\rightarrow \nu_s$ conversion due to the damping term $D(p)=\Gamma_{\alpha}/2$  in the denominator of Eq.\ (\ref{eq:Boltzmann}), leading to larger mixing angles being needed to produce the observed relic density of DM. However, as $\epsilon$ increases, the rate at which active neutrinos are produced in the early Universe through the new interaction also increases. This ends up producing the observed relic for  smaller mixing angles. The suppressing power of NSSI decreases in presence of heavier mediators, and for $m_\phi=100$ GeV the values of allowed $\sin^2(2\theta)_{\text{NSSI}}$ are mostly less than $\sin^2(2\theta)_{\text{DW}}$.

The expected sensitivities of upcoming experiments are also reported on the plots where the sensitivity regions are enclosed by blue lines for ECHo \cite{Gastaldo:2017edk}, black and gray lines for TRISTAN \cite{KATRIN:2018oow}, and beige lines for HUNTER \cite{Martoff:2021vxp}. 
We do not report in the plots any line to represent the sensitivity of the BeEST experiment \cite{Fretwell:2020ntq} because at the moment it is too limited to be important for our study and it is strongly reduced by the constraint deriving from the requirement for the DM candidate to be stable on timescales comparable with the age of the Universe (gray region in the top right corner of the plots).
While TRISTAN's and ECHo's sensitivities are way too far from the region interested by the effects of NSSI, the phase 3 of the HUNTER experiment will probe masses and mixing angles for which the entire abundance of DM could be constituted by sterile neutrinos for $\epsilon_{P, A} \lesssim 10$.
In the case of light NSSI mediator ($m_\phi = 10$ GeV), the interesting region of the parameter space probed by HUNTER will be larger, but some values of $m_s$, $\sin^2(2\theta)$, and $\epsilon_{P, A}$ will be probed also in case of heavier mediators ($m_\phi = 50, 100$ GeV).

We have furthermore added four benchmark points:  
\begin{itemize}
    \item BP1 corresponds to the observed X-ray monochromatic line at 3.55 keV \cite{Bulbul:2014sua,Boyarsky:2014jta}, only in the case of $\Omega_s = \Omega_{\text{DM}}$;
    \item BP2 is characterized by corresponding to the smallest mixing angle that will be probed by HUNTER and, at the same time, by fulfilling the condition $\Omega_s = \Omega_{\text{DM}}$ in the presence of NSSI with all three mediator mass values considered here;
    \item BP3 is characterized by the largest $m_s$ value not constrained by any phase space argument and, at the same time, by fulfilling both the conditions $\Omega_s = \Omega_{\text{DM}}$ and $\Omega_s = 0.1 \times \Omega_{\text{DM}}$ in different NSSI scenarios;
    \item BP4 is characterized by a mass that makes it  an almost ``cold'' DM candidate in the standard DW scenario.
\end{itemize}

In what regards laboratory constraints on NSSI, only loose limits exist. In particular, for heavy mediator masses as we consider here, the usual low energy limits from meson decays or double beta  decays, see e.g.\ \cite{Blinov:2019gcj,Deppisch:2020sqh}, do not apply. The same holds for astrophysical constraints such as from BBN or supernova considerations.  
The only relevant limits for the case presented in this work come from  $Z$-boson decay at one loop level \cite{Bilenky:1999dn}. Using only vector NSSI among Dirac neutrinos, the authors showed that the constraints on NSSI couplings can be as strong as $|\epsilon| \lesssim 2$. However, there exists a possibility of cancellation of NSSI couplings among different neutrino flavors, and this can loosen the bound to $|\epsilon|\lesssim 250$. While these bounds cannot be directly applied to scalar, pseudoscalar or axial-vector NSSIs among Majorana neutrinos, one can translate these bounds using Fierz rearrangement. We do not explicitly calculate the bounds, rather restrict our NSSI couplings to be $|\epsilon_{S,P,A}|\lesssim 100$.

The biggest restriction on sterile neutrino dark matter, also in the case of production assisted by NSSI, is represented by the constraint coming from X-ray observations \cite{Perez:2016tcq, Ng:2019gch}, represented by the purple line in the plots. We note that those limits could be evaded for instance by additional decay modes of the sterile neutrinos, see e.g.\ \cite{Barry:2014ika,Benso:2019jog}. \\

\subsection{Cocktail dark matter scenario}
\noindent
In the right panels  of FIGS.\ \ref{fig:pseudoscalar} and  \ref{fig:axial-vector} we summarize the results obtained for the case of  ``Cocktail dark matter'', with the same benchmark points as the previous section.
The hypothesis here is that only a fraction of dark matter (10\% in the plots we show) is constituted by sterile neutrinos, while about the rest of the DM abundance we remain agnostic. 
In this case, we observe a shift towards smaller values of mixing angle of the band relative to the production assisted by NSSI. 
The sensitivity region of upcoming experiments is unvaried by this assumption, since it depends only on the value of the mixing angle in the vacuum and not on the abundance of sterile neutrinos. 
This implies that the parameter space region in which the condition $\Omega_{s} h^2 = 0.1 \times \Omega_{\text{DM}} h^2 = 0.012$ is verified will not be testable even by HUNTER.
On the other hand, the constraint from X-ray observations is visibly relaxed in this scenario, because of its dependence on the sterile neutrino abundance, as already discussed e.g.\ in \cite{Benso:2019jog}. 
Thus, in this case, even larger ranges of values of $m_s$ and $\sin^2(2\theta)$ are allowed from this point of view.
A further advantage of the ``Cocktail DM'' scenario is that the remaining fraction of DM might have completely different features from those of sterile neutrinos, and the mixture of the two or more candidates could fit much easier also all the constraints coming from other observables, such as for example from structure formation.

\subsection{Impact on structure formation}\label{sec:struc}
\noindent
Following \cite{Merle:2015vzu}, it is possible to get a semi-analytical solution for the Boltzmann equation reported in Eq.\ (\ref{eq:Boltzmann_T}). The distribution function of sterile neutrinos is
\begin{widetext}
\begin{equation}
    f_{s}(p_f, T_f) = \int_{T_i}^{T_f} - \frac{1}{H T} \left[ 1 + \frac{1}{3} \frac{T g'_S}{g_S} \right] \cdot h \left( p(T_f) \frac{T}{T_f} \left(\frac{g_S(T)}{g_S(T_f)}\right)^{1/3}, T \right) \cdot f_{\alpha}\left( p(T_f) \frac{T}{T_f} \left(\frac{g_S(T)}{g_S(T_f)}\right)^{1/3}, T \right) dT\,,
    \label{eq:solution}
\end{equation}
\end{widetext}
where the function $h$ contains the details of the Dodelson-Widrow mechanism, in our case modified by NSSI.
For simplicity, in this section we focus only on pseudoscalar NSSI, but analogous discussion and conclusions hold for scalar and axial-vector NSSI. 
The distribution function is a relevant quantity if we want to confront our results with the  constraints coming from structure formation, and particularly important is the high momentum part of the distribution. 
It is often reported that the distribution function of sterile neutrinos produced through Dodelson-Widrow mechanism can be approximated by a suppressed thermal Fermi-Dirac distribution.
However, this is true only under certain conditions. In particular, the number of relativistic degrees of freedom $g_S$ should vary so slowly with $T$ that it is possible to replace it by its average value $\langle g_S\rangle$ and pull the thermal part $f_\alpha$ in front of the integral. Moreover, $h$ would need to vary only very slowly with  momentum $p$.

\begin{figure}[t!]
	\centerline{\includegraphics[width=0.45\textwidth]{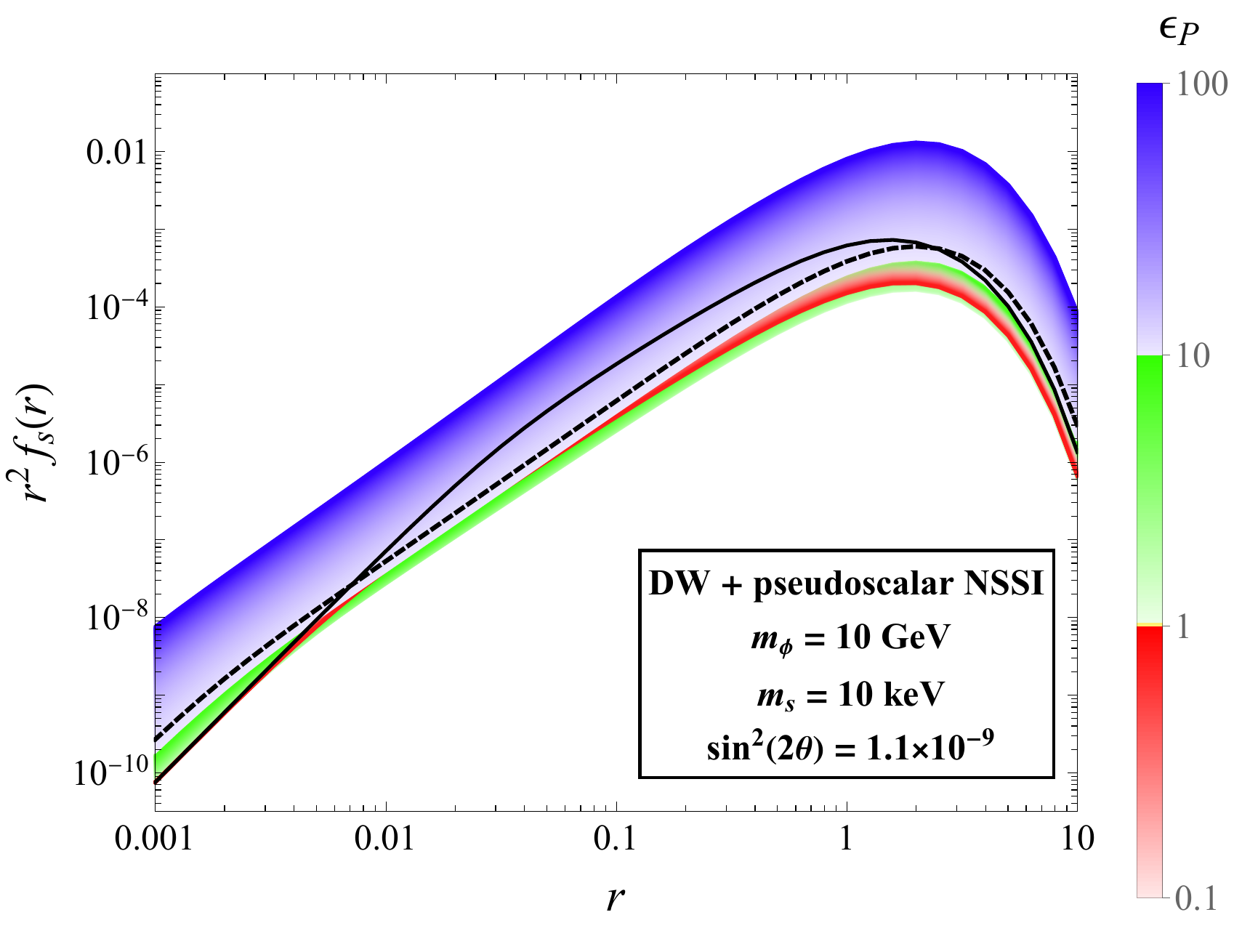}}
	\caption{\small Sterile neutrino momentum distribution $r^2 f(r)$ where $r=p/T$ in the case of Dodelson-Widrow production modified by the action of pseudoscalar NSSI. The values of the sterile neutrino parameters are chosen as $m_s = 10$ keV, $\sin^2(2\theta) = 1.1 \times 10^{-9}$ (BP2 in FIGS.\ \ref{fig:pseudoscalar} and \ref{fig:axial-vector}). In different shades of red, green, and blue, we observe the result corresponding to different values of $\epsilon_P$. The black dashed line represents the distribution function that gives $\Omega_s\,h^2= \Omega_{\text{DM}}\,h^2= 0.12$. The black solid line represents the standard Dodelson-Widrow case where the pseudoscalar NSSI is switched off.}
	\label{fig:r2f}
\end{figure}

In the case we consider, where NSSI are involved in the production of sterile neutrino dark matter, the shape of the distribution function is further modified by the action of such NSSI with respect to the distribution function obtained in the standard Dodelson-Widrow scenario. 
This is shown in FIG. \ref{fig:r2f}, where we plot the momentum distribution function $r^2 f(r)$, where $r = p/T$, for the case of the lightest pseudoscalar NSSI mediator $m_\phi = 10$ GeV, $m_s = 10$ keV, $\sin^2(2\theta) = 1.1 \times 10^{-9}$ and varying strength of the NSSI parametrized by $\epsilon_P$. 
The thick black line represents the result in the standard Dodelson-Widrow scenario. 
The dashed black line represents the result corresponding to the value of $\epsilon_P$ that, with the chosen values of $m_s$ and $\sin^2(2\theta)$, produces $\Omega_s\,h^2= \Omega_{\text{DM}}\,h^2= 0.12$.

\begin{figure}[t!]
	\centerline{\includegraphics[width=0.45\textwidth]{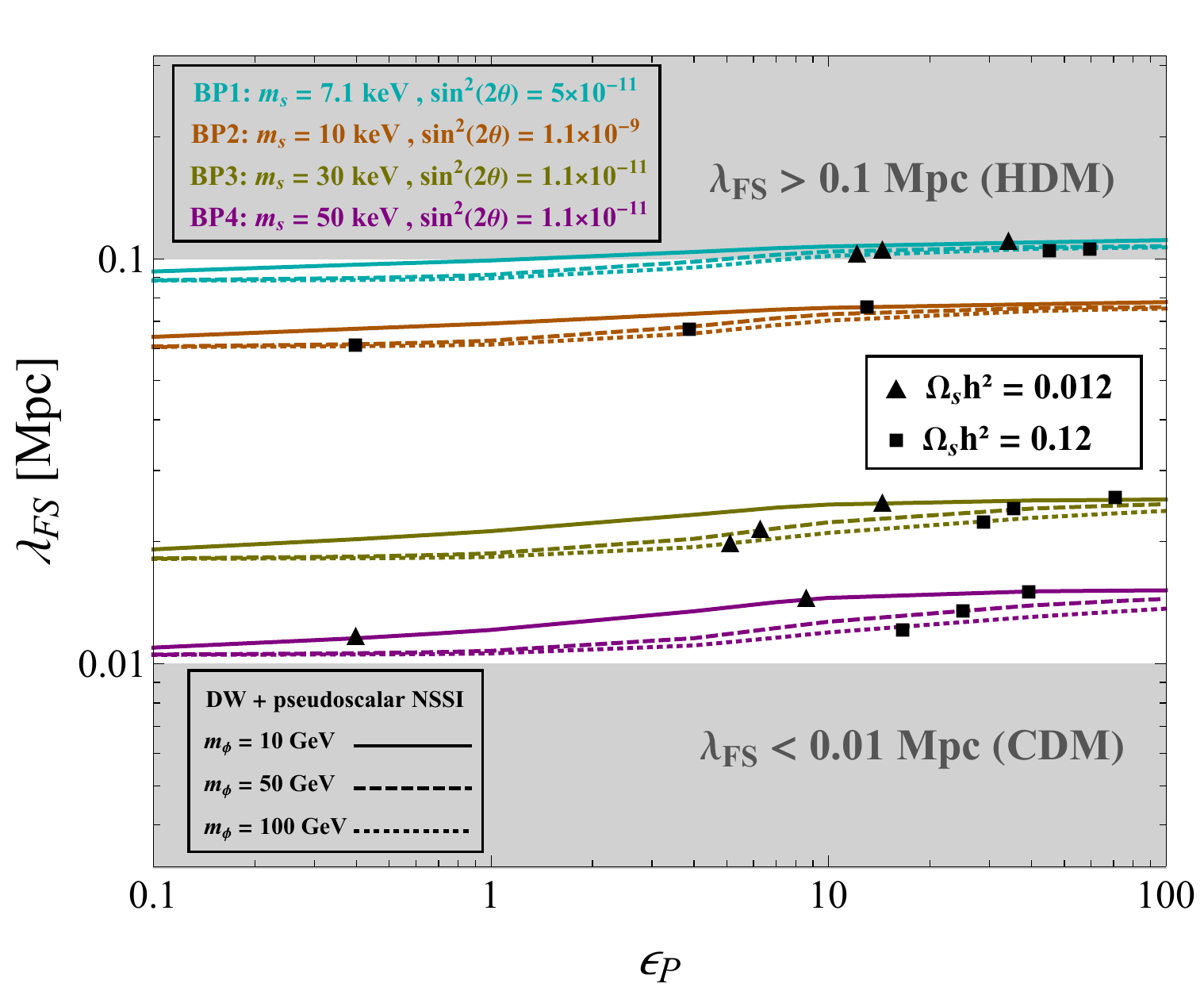}}
	\caption{\small Variation of the free streaming length of sterile neutrino dark matter determined by the increasing strength of NSSI for different values of $m_s$ and $\sin^2(2\theta)$. Each color refers to a benchmark point given in FIGS.\ \ref{fig:pseudoscalar} and \ref{fig:axial-vector}. Each line type corresponds to a different value of the NSSI mediator mass. Black squares pinpoint to values of $\epsilon_P$ for which the condition $\Omega_s\,h^2= \Omega_{\text{DM}}\,h^2= 0.12$ is satisfied. Black triangles identify values of $\epsilon_P$ such that only the 10\% of the DM abundance is constituted by sterile neutrinos in the ``cocktail DM'' scenario.}
	\label{fig:lfs}
\end{figure}

The impact of DM on structure formation can be estimated through the calculation of the free streaming length \cite{Palazzo:2007gz}
\begin{equation}
 \lambda_{\text{FS}} = \int_{0}^{t_0} \frac{\langle v(t) \rangle}{a(t)} dt \simeq 1.2 \text{ Mpc} \left( \frac{\text{keV}}{m_s} \right) \frac{\langle p/T \rangle}{3.15},
\end{equation}
where $t_0$ is the time today, because the largest scale affected by free-streaming is nothing but the present value of the particle horizon of warm particles with a typical velocity $\langle v(t) \rangle$ \cite{Boyarsky:2008xj}. 
In the approximated form, we see that the value of the free-streaming length depends on the features of the production through the distribution function used to obtain the typical value of $p/T$. 
For a DM candidate to be considered ``warm'', and thus compatible with the structures that we observe in today's Universe, the free-streaming length must be $0.01~\text{Mpc}	< \lambda_{\text{FS}} < 0.1~\text{Mpc}$ \cite{Merle:2013wta}. This is what happens for the majority of the cases that we consider, see FIG. \ref{fig:lfs}. Moreover, we notice that no effect of NSSI in the allowed range of values of $\epsilon$ is strong enough to drastically modify the impact of sterile neutrino dark matter on structure formation, from the ``cold'' regime to the ``hot'' one, or vice versa.

For our four benchmark points, we show in FIG.\ \ref{fig:lfs} the impact of pseudoscalar NSSI on the free streaming length, varying NSSI strengths $\epsilon_P$ and mediator masses. 
Different colors correspond to different benchmark points, while different line types are related to different mass values of the mediators. 
Black squares identify values of $\epsilon_P$ that give $\Omega_s\,h^2= \Omega_{\text{DM}}\,h^2= 0.12$ for the chosen values of $m_s$ and $\sin^2(2\theta)$, while black triangles pinpoint values of $\epsilon_P$ that give $\Omega_s\,h^2= 0.1 \times \Omega_{\text{DM}}\,h^2= 0.012$. 

The lines corresponding to the chosen benchmark points span the entire region in which sterile neutrinos can be considered warm DM candidates. 
Their location is determined by the mass of the sterile neutrinos rather than their interactions. 
We notice also that the influence of NSSI on the free streaming length is limited and the impact is small even for large values of $\epsilon_P$. 
This allows us to say that the existence NSSI acting with strengths $\epsilon_P$ within the current limits, would not put sterile neutrinos in tension with structure formation constraints, unless they are very light. 
On the other hand, if we consider BP1 that identifies the famous observed X-ray line at 3.55 keV \cite{Bulbul:2014sua,Boyarsky:2014jta}, we see that large NSSI would be needed to produce an abundance of such sterile neutrinos large enough to constitute a non negligible percentage of the Universe's DM content.
However, such  large NSSI would put sterile neutrinos with such features in conflict with constraints coming from structure formation: they would have been produced with too high velocities modifying large structures that we observe
today.

BP2 is particularly interesting. It represents a case in which the NSSI effect is crucial to allow sterile neutrinos to be produced in the correct abundance and at the same time it does not lead to tensions with structure formation. 
Moreover, being at the border of the sensitivity region expected for the phase 3 of the HUNTER experiment, the values of the parameters relative to this point will be available for experimental test.

\subsection{Comment on the truncation at the second order}
\begin{figure}[!t]
	\centerline{\includegraphics[width=0.45\textwidth]{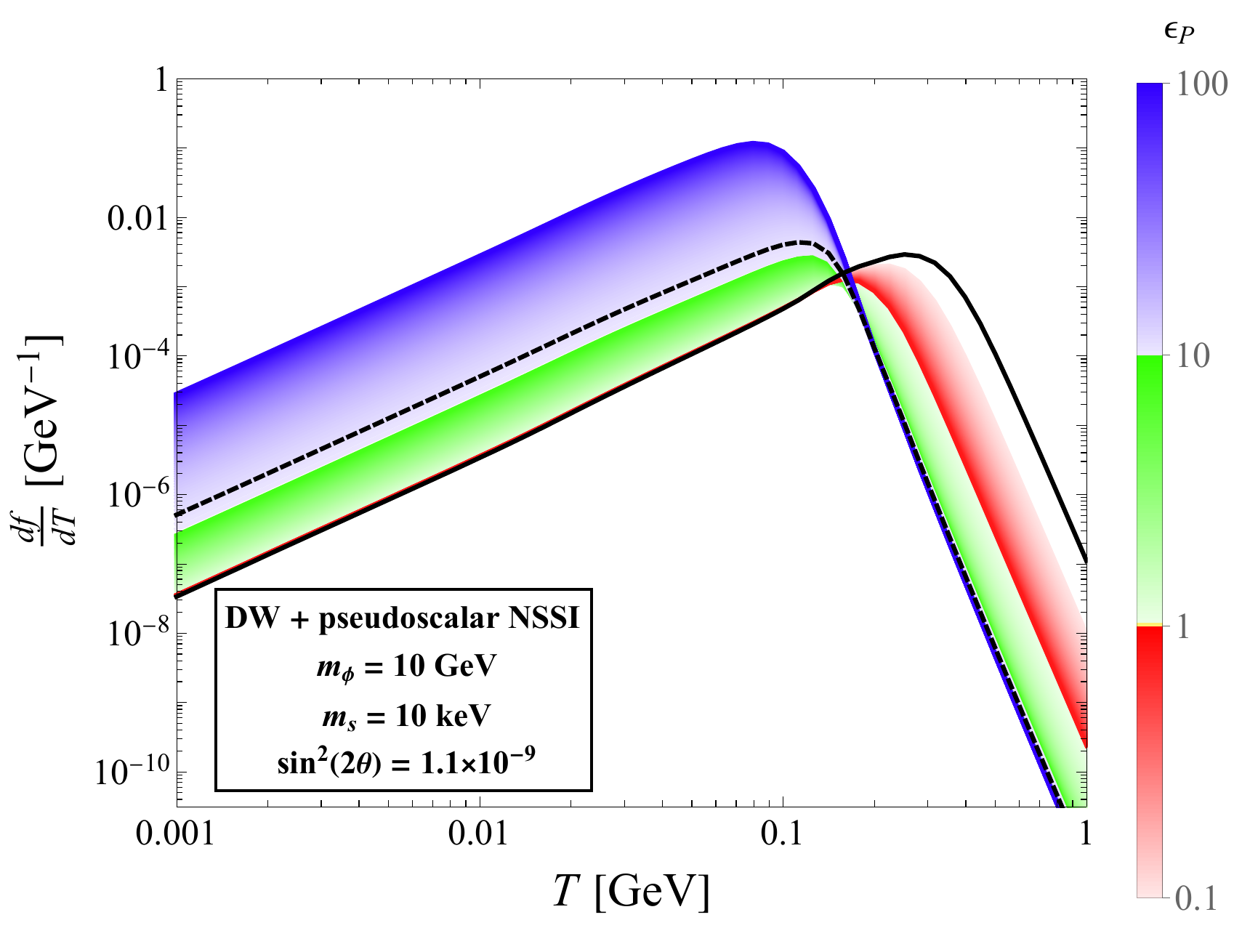}}
	\caption{\textmd{ \small Evolution of the production rate of sterile neutrino with $m_s = 10$ keV and $\sin^2(2\theta) = 1.1 \times 10^{-9}$ (BP2 in FIG.\  \ref{fig:pseudoscalar}) with temperature. The black thick line corresponds to the standard Dodelson-Widrow production case. The black dashed line corresponds to the production assisted by pseudoscalar NSSI with $\epsilon_P$ such that $\Omega_s\,h^2= \Omega_{\text{DM}}\,h^2= 0.12$. Different shades of red, green and blue correspond to increasing strength of NSSI involved in the sterile neutrino production. All the lines are obtained under the hypothesis that the NSSI mediator has mass $m_\phi = 10$ GeV.} 
	}
	\label{fig:dfdT}
\end{figure}
\noindent
In FIG.\ \ref{fig:dfdT} we see the evolution of the production rate of sterile neutrinos with $m_s = 10$ keV and $\sin^2(2\theta) = 1.1 \times 10^{-9}$ produced while the Universe cools down from 1 GeV to 1 MeV.
In the standard Dodelson-Widrow case (black thick line), the peak of the production occurs between $T \sim 200$ MeV and $T \sim 300$ MeV in agreement with the approximate expression $133 \, (m_s/{\rm keV})^\frac 13 \,{\rm MeV}$. 
The presence of the pseudoscalar NSSI with light ($m_\phi = 10$ GeV) mediator modifies the mechanism and shifts the peak of the production towards lower temperatures. As shown by the progression of the colors in the plot, the larger the strength of NSSI, the lower the peak temperature. 
In particular, the abundance of sterile neutrinos sufficient to constitute the entire content of DM in the Universe, corresponds to the value of $\epsilon_P$ represented by the black dashed line whose peak is at $T \sim 100$ MeV. 
This temperature is much lower than the value of the NSSI mediator, and thus our choice of treating the impact of NSSI in an effective framework with truncation at the second order in the expansion in $p$ is justified.

\section{Conclusions and Outlook}
\label{sec:conclusion}
\noindent 
Sterile neutrinos are popular dark matter candidates. In the near future laboratory limits on such particles will improve considerably. 
We have investigated here how the presence of neutrino self-interactions would influence the parameter space of mass and mixing of keV-scale sterile neutrinos. We have shown that the parameter space widens, and can move in the direction of the upcoming experiment, in particular HUNTER. We have illustrated that a meaningful  EFT analysis needs to include momentum-dependent terms, in order to have the physically correct temperature dependence of the thermal potential. The free-streaming length of sterile neutrino dark matter is influenced by the presence of neutrino self-interactions, though it stays within the ``warm values'' between 0.01 ad 0.1 Mpc. 

One can think of several extensions of our study, applying it to Dirac neutrinos, other production mechanisms of sterile neutrino dark matter, additional interactions of the sterile neutrinos or of other particles in the thermal plasma such as electrons. Given the theoretical motivation of such frameworks, and the amount of several experimental approaches to keV-scale neutrinos, such studies are surely worthwhile. 

\section*{Acknowledgments}
\label{sec:Acknowledgments}
\noindent 
We thank Evgeny Akhmedov, Ingolf Bischer, Irene Tamborra, Yue Zhang and Walter Tangarife for helpful discussions. We are grateful to Johannes Herms for having pointed out an inaccuracy about the X-ray bound relaxation present in the previous version of the paper. CB acknowledges support of the  IMPRS-PTFS.

\bibliography{references}

\onecolumngrid
\appendix
\renewcommand{\thesection}{A}
\renewcommand{\theequation}{A.\arabic{equation}}
\section{Appendix: NSSI Thermal Potential}
\label{appendix:potential}
\noindent
In this Appendix we will calculate the effective potential of our scenario, in order to show how the temperature dependence arises in our treatment of effective operators. The rates can be calculated in analogy. 

Following the treatment provided in \cite{DOlivo:1992lwg} for the Standard Model thermal potential calculations, we calculate the NSSI contribution to effective potential of active neutrinos in early Universe. It is essential to use higher order terms in the effective NSSI Lagrangian to capture the momentum-dependence of the thermal potential.

Starting with a Yukawa-like interaction between active neutrinos $\nu$ and a complex scalar mediator  $\phi$ (for pseudoscalar or axial-vector particles the calculation  is similar), the full Lagrangian can be written as
\eqn{
\label{eqn:lagrangianfull}
\mathcal{L}\lo{full}= \partial_\mu \phi^\dagger \partial^\mu\phi-m^2 \phi^\dagger \phi +\lambda_{\phi} \bar{\nu} \mathcal{O} \nu \phi +\lambda_{\phi}^\ast \phi^\dagger \bar{\nu} \bar{\mathcal{O}} \nu  \ , \quad \text{where } \bar{\mathcal{O}}=\gamma^0\mathcal{O}^\dagger \gamma^0\ ,
}
where $\mathcal{O}$ is an element of a complete set of bilinear covariants $\{\mathbb{I}, \gamma^\mu, i\gamma^5, \gamma^\mu\gamma^5 , \sigma^{\mu\nu}\}$. 
However, if the mediator is much heavier than the temperature range we are interested in ($m_\phi \gg T$),  we can employ an effective field theory framework and integrate out heavy degrees of freedom in the full Lagrangian. To find the EFT Lagrangian, we first solve equation of motion for the heavy degree of freedom $\phi$,
\eqn{
\pdv{\mathcal{L}\lo{full}}{\phi} - \partial_\mu \pdv{\mathcal{L}\lo{full}}{(\partial_\mu{\phi})}=0\ .
}
Solving the equation of motion
\eqn{
-m^2\phi^\dagger +\lambda_{\phi} \bar{\nu} \mathcal{O} \nu -\partial_\mu  \partial^\mu \phi^\dagger &=0 \Rightarrow
(\Box + m^2)\phi^\dagger= \lambda_{\phi} \bar{\nu} \mathcal{O} \nu \ ,
}
and we obtain an expression for $\phi^\dagger$,
\eqn{
\phi^\dagger= \frac{\lambda_{\phi} \bar{\nu}\mathcal{O} \nu}{(\Box + m^2)} \ .
}
A similar expression,
\eqn{
 \phi= \frac{\lambda_{\phi}^\ast \bar{\nu} \bar{\mathcal{O}} \nu}{(\Box + m^2)} \ ,
}
can be obtained by solving equation of motion for $\phi^\dagger$.
Substituting these in the full Lagrangian Eq.~\eqref{eqn:lagrangianfull} to integrate out the heavy complex scalar $\phi$, we obtain
\eqn{
\mathcal{L}\lo{NSSI} = \lambda_{\phi}^2\frac{(\bar{\nu} \mathcal{O} \nu )(\bar{\nu} \bar{\mathcal{O}} \nu)}{\left(\Box + m_\phi^2\right)} . 
}
Keeping terms up to first order in $\Box$ to retain momentum dependence, we have
\eqn{
\mathcal{L}\lo{NSSI} = \frac{G_{\phi}}{\sqrt{2}}\left( (\bar{\nu} \mathcal{O} \nu )(\bar{\nu} \bar{\mathcal{O}} \nu) - (\bar{\nu} \mathcal{O} \nu )\frac{\Box}{m_\phi^2}(\bar{\nu} \bar{\mathcal{O}} \nu)\right)  , \text{ where } G_{\phi}=  \frac{\sqrt{2}\lambda_{\phi}^2}{m_\phi^2}
}
is the strength of NSSI defined similar to the Fermi constant $G_F$.

Using $G_{\phi}= G_F \epsilon$, where $\epsilon$ indicates the NSSI strength compared to the standard weak interactions, we get the final form of NSSI Lagrangian:
\eqn{
\label{eqn:hoLagrangian}
\mathcal{L}\lo{NSSI} = \frac{G_F \epsilon }{\sqrt{2}} (\bar{\nu} \mathcal{O} \nu )(\bar{\nu} \bar{\mathcal{O}} \nu) -\frac{G_F \epsilon }{\sqrt{2}} (\bar{\nu} \mathcal{O} \nu )\frac{\Box}{m_\phi^2}(\bar{\nu} \bar{\mathcal{O}} \nu) \ ,
}
with $\mathcal{O}=\mathbb{I}, i\gamma^5, \gamma^\mu \gamma^5$ giving scalar, pseudoscalar and axial-vector NSSI, respectively. 
The second term in the Lagrangian Eq.~\eqref{eqn:hoLagrangian} gives momentum-dependent Feynman rules and is essential to capture the temperature dependence of thermal 
potential. For a vector propagator there will be a relative minus sign in the effective Lagrangian  which comes from the fact that scalar propagators $\Delta_\phi =\frac{i}{q^2-m_\phi^2}$ have an extra minus sign compared to  vector propagators $\Delta_V =\frac{-ig^{\mu\nu}}{q^2-m_V^2}$. 

\subsection{Calculation of $\Sigma\lo{eff}$}
\noindent 
The NSSI contribution to the self-energy for scalar and pseudoscalar mediators (see FIG.\ \ref{fig:PotentialDiagrams}) 
can be written as
\begin{equation}
      i\Sigma\lo{eff} = \frac{G_{F}{\epsilon}_{j}}{\sqrt{2}}\int \frac{d^4 p }{(2\pi)^4}(\mathcal{O}_{j} + C\mathcal{O}_{j}^\text{T}C^{-1})S_{F}(p)(\bar{\mathcal{O}}_{j}+ C{\bar{\mathcal{O}}_{j}}^\text{T}C^{-1}) \bigg(1+\frac{q_\mu q^\mu}{m_\phi^2} \bigg)\ , 
\end{equation}
where $q_\mu=k_\mu - p_\mu$  denotes the difference between the neutrino momentum $k$ and four-velocity of medium $u$. The second term $q^2/m_\phi^2$ in the final brackets comes from the second term in the Lagrangian Eq.~\eqref{eqn:hoLagrangian} and eventually lead to a temperature dependence in the potential. 
The extra factors with charge conjugation operator come from the Majorana Feynman rules~\cite{Pal:2010ih}.

In the finite-temperature field theory  formalism, the fermion propagator $S_F(p)$ is given as
\begin{equation}
S_{F}\left( p\right) =\left( \slashed{p}+m_l\right) \left[ \frac{1}{p^{2}-m^{2}_l}+2\pi i\delta \left( p^{2}-m^{2}_l\right)\eta \left( p\cdot u\right)\right] ,
\end{equation}
where
\begin{equation}
\eta \left( p\cdot u\right) =\frac{\theta \left( p\cdot u\right) }{e^{x}+1}+
\dfrac{\theta \left( -p\cdot u\right) }{e^{-x}+1},\,  x=\dfrac{\left( p\cdot u-\mu \right) }{T} \, ;
\end{equation}
$\theta(x)$ is the unit step function and $n_{\pm}= \left(e^{\pm x}+1\right)^{-1}$ is the occupation number for the background fermions ($n_{+}$) and antifermions ($n_{-}$), in principle containing a chemical potential $\mu$.

The background-independent part of the fermion propagator only renormalizes the wave function and does not contribute to the dispersion relation in the lowest order \cite{DOlivo:1992lwg}. To simplify the calculations, we will only consider background-dependent 
\begin{equation}
   S_{F}^{T}\left( p\right) =2\pi i \delta \left( p^{2}-m^{2}_l\right)\eta \left( p\cdot u\right)\left( \slashed{p}+m_l\right),
\end{equation}
where the superscript $T$ indicates the temperature dependence of fermion propagator.\\
Using $q_\mu q^\mu = (p-k)^2 \approx -2\,p\cdot k$, temperature dependent self-energy,
\eqn{
 i\Sigma^{(T)}\lo{eff} =\frac{ G_{F}{\epsilon}_{j}}{\sqrt{2}}\int \frac{d^4 p }{(2\pi)^4}(\mathcal{O}_{j} + C\mathcal{O}_{j}^\text{T}C^{-1})S_{F}^{T}\left( p\right)(\bar{\mathcal{O}}_{j}+ C{\bar{\mathcal{O}}_{j}}^\text{T}C^{-1})\bigg(1-\frac{2p\cdot k}{m_\phi^2} \bigg) \,.
}

The effective potential $\mathcal{V}_\alpha$ in the lowest order is given by neutrino dispersion relations as $\mathcal{V}_\alpha = b_0(\omega_k=\kappa)$ for $\nu$ and $\mathcal{V}_\alpha = -b_0(\omega_k=-\kappa)$ for $\bar{\nu}$ \cite{DOlivo:1992lwg}, where $b_0$ can be calculated from $\Sigma\lo{eff} = a_0 \slashed{k}+ b_0 \slashed{u}$.

For scalar NSSI with $\mathcal{O}_{S} = \mathbb{I}$, the self-energy becomes 
\begin{equation}\label{scalarvt}
    \begin{split}
       i\Sigma^{(T)}\lo{eff} &= \frac{G_{F}{\epsilon}_{S}}{\sqrt{2}}  \int \frac{d^4 p }{(2\pi)^4}(\mathcal{O}_{S}+ C{\mathcal{O}_{S}}^\text{T}C^{-1})S_{F}^{T}\left( p\right)(\bar{\mathcal{O}_{S}}+ C{\bar{\mathcal{O}_{S}}}^\text{T}C^{-1})\bigg(1-\frac{2p\cdot k}{m_\phi^2} \bigg)\\
       &=\frac{ 4G_{F}{\epsilon}_{S}}{\sqrt{2}}   \int \frac{d^4 p }{(2\pi)^4} S_{F}^{T}(p)\bigg(1-\frac{2p\cdot k}{m_\phi^2} \bigg)\\
        &=\frac{ 4G_{F}{\epsilon}_{S}}{\sqrt{2}}   \int \frac{d^4 p }{(2\pi)^4}\cdot 2\pi i \delta \left( p^{2}-m^{2}_\nu\right)\eta \left( p\cdot u\right)  \gamma_\mu p^\mu \cdot \bigg(1-\frac{2p\cdot k}{m_\phi^2} \bigg) \\
       &= \frac{4iG_{F}{\epsilon}_{S}}{\sqrt{2}}   \int \frac{d^4 p }{(2\pi)^3} \delta \left( p^{2}-m^{2}_\nu\right)\eta \left( p\cdot u\right)  \gamma_\mu p^\mu \bigg(1-\frac{2p\cdot k}{m_\phi^2} \bigg)\ .
    \end{split}
\end{equation}
Defining two momentum dependent integrals
\eqn{
I_{\mu}&=\int \frac{d^{4} p}{(2 \pi)^{3}} \delta\left(p^{2}-m_{\nu}^{2}\right) \eta(p \cdot u) p_{\mu}\ ,\\
I_{\mu\nu}&=\int \frac{d^{4} p}{(2 \pi)^{3}} \delta\left(p^{2}-m_{\nu}^{2}\right) \eta(p \cdot u) p_{\mu} p_{\nu}\ ,
}
we can rewrite Eq.~\eqref{scalarvt} as
\begin{equation}\label{s3}
    \begin{split}
      i\Sigma^{(T)}\lo{eff}= \frac{4iG_{F}{\epsilon}_{S}}{\sqrt{2}}   \gamma^\mu\left[ I_\mu- \frac{2 k^\nu }{m_\phi^2} I_{\mu\nu}\right].
    \end{split}
\end{equation}
We can calculate the integral $I_{\mu}=A u_\mu $, which is manifestly covariant and has only $u_{\mu}$ dependence, by contracting it with $u^\mu$.
\begin{equation}
 I_{\mu}u^\mu=A u_\mu u^\mu = A =\int \frac{d^{4} p}{(2 \pi)^{3}} \delta\left(p^{2}-m_{\nu}^{2}\right) \eta(p \cdot u) p_{\mu}u^\mu= J_{1}^{(\nu)}\ .
\end{equation}
Where $J_{n}^{(f)}$ is defined as $J_{n}^{(f)}=\int \frac{d^{4} p}{(2 \pi)^{3}} \delta\left(p^{2}-m_{f}^{2}\right) \eta(p \cdot u)(p \cdot u)^{n}$ and calculated below. 

Now we have $    I_{\mu}= J_{1}^{(\nu)}u_\mu$. 
Similarly, $I_{\mu\nu}$ can be obtained by contracting $I_{\mu \nu}=A g_{\mu \nu}+B u_{\mu} u_{\nu}$ with $u_{\mu} u_{v}$ and $g_{\mu \nu}$. One finds 
\begin{equation}
 g^{\mu \nu}I_{\mu \nu}=\int \frac{d^{4} p}{(2 \pi)^{3}} \delta\left(p^{2}-m_{\nu}^{2}\right) \eta(p \cdot u) p_{\mu}  g^{\mu \nu}p_{\nu} =m_{\nu}^{2} J_{0}^{(\nu)}=4A+B\ , 
\end{equation}
\begin{equation}
u^{\mu} u^{\nu}I_{\mu \nu}=\int \frac{d^{4} p}{(2 \pi)^{3}} \delta\left(p^{2}-m_{\nu}^{2}\right) \eta(p \cdot u) p_{\mu} u^{\mu} u^{\nu}p_{\nu} = J_{2}^{(\nu)}=A+B  \ .
\end{equation}
Solving for $A$ and $B$, we get
\eqn{
A&=\frac{1}{3}\left(m_{\nu}^{2} J_{0}^{(\nu)}-J_{2}^{(\nu)}\right), \\
B&=\frac{1}{3}\left(4 J_{2}^{(\nu)}-m_{\nu}^{2} J_{0}^{(\nu)}\right) .
}
Substituting these expressions for $I_{\mu}$ and $I_{\mu v}$ into Eq.\  \eqref{s3},
\eqn{
     i\Sigma^{(T)}\lo{eff}  & = \frac{4iG_{F}{\epsilon}_{S}}{\sqrt{2}}   \left[ J_{1}^{(\nu)}\slashed{u}- \frac{2 \gamma^\mu k^\nu }{3m_\phi^2} \left(m_{\nu}^{2} J_{0}^{(\nu)}-J_{2}^{(\nu)}\right) g_{\mu \nu}-\frac{2 \gamma^\mu k^\nu }{3m_\phi^2}\left(4 J_{2}^{(\nu)}-m_{\nu}^{2} J_{0}^{(\nu)}\right) u_{\mu} u_{\nu}\right]\\
     &= \frac{4iG_{F}{\epsilon}_{S}}{\sqrt{2}}   \left[ J_{1}^{(\nu)}\slashed{u} -\frac{2 }{3m_\phi^2} \left(m_{\nu}^{2} J_{0}^{(\nu)}-J_{2}^{(\nu)}\right) \slashed{k}-\frac{2 \omega }{3m_\phi^2}\left(4 J_{2}^{(\nu)}-m_{\nu}^{2} J_{0}^{(\nu)}\right) \slashed{u}\right] .
     }
Taking the neutrino mass $m_\nu \approx 0$, this simplifies to 
\begin{equation}\label{s2}
     i\Sigma^{(T)}\lo{eff}  = \frac{4iG_{F}{\epsilon}_{S}}{\sqrt{2}}   \left[-\frac{2 J_{2}^{(\nu) }}{3m_\phi^2}  \slashed{k}+ \bigg(J_{1}^{(\nu)} -\frac{8 \omega }{3m_\phi^2} J_{2}^{( \nu)}\bigg) \slashed{u}\right]\ .
\end{equation}
Now comparing Eq.\  \eqref{s2}  with $\Sigma^{(T)}\lo{eff} = a_0 \slashed{k}+ b_0 \slashed{u}$, one finds 
\begin{equation}
    b_0 =\frac{4G_{F}{\epsilon}_{S}}{\sqrt{2}}  \left( J_{1}^{(\nu)} -\frac{8 \omega }{3m_\phi^2} J_{2}^{( \nu)}\right) .
\end{equation}
From the calculation of $J_n^{(\nu)}$ provided below, the result is 
\begin{equation}
    b_0 =\frac{4G_{F}}{\sqrt{2}}{\epsilon}_{S} \left[\frac{1}{2}\left(n_{\nu}-n_{\bar{\nu}}\right)-\frac{8 \omega}{3 m_\phi^2}\cdot \frac{1}{2}\left(n_{\nu}\left\langle E_{\nu}\right\rangle+n_{\bar{\nu}}\left\langle E_{\bar{\nu}}\right\rangle\right)\right] . 
\end{equation}
The first term becomes zero if we assume a lepton symmetric Universe. In this case the  scalar NSSI thermal potential at the lowest order in $\omega$ is 
\begin{equation}
    \mathcal{V}_S=  -\frac{7 \sqrt{2}\pi ^2G_{F}{\epsilon}_{S} }{45m_\phi^2}\cdot \omega T^4\ .
\end{equation}
Similarly for an axial-vector NSSI with $\Gamma_{A} = \gamma_\mu\gamma^5$, taking the extra minus sign of the effective Lagrangian into account, we find  
\begin{equation}
    \mathcal{V}_A=  -\frac{14 \sqrt{2}\pi ^2 G_{F}{\epsilon}_{A}}{45m_\phi^2}\cdot \omega T^4\ ,
\end{equation}
and a pseudoscalar NSSI with $\Gamma_{P} = i\gamma^5$,
\begin{equation}
    \mathcal{V}_P=  -\frac{7 \sqrt{2}\pi ^2G_{F}{\epsilon}_{P} }{45m_\phi^2}\cdot \omega T^4\ .
\end{equation}
\subsection{Evaluating $J_{n}^{(f)}$}
\label{subsec:jn}
\noindent 
Our task here is to evaluate 
\begin{equation}
  J_{n}^{(f)}=\int \frac{d^{4} p}{(2 \pi)^{3}} \delta\left(p^{2}-m_{f}^{2}\right) \eta(p \cdot u)(p \cdot u)^{n}\ ,
\end{equation}
with 
\begin{equation}
\eta \left( p\cdot u\right) =\frac{\theta \left( p\cdot u \right) }{\exp\left({\frac{p\cdot u - \mu}{T}}\right)+1}  +
\frac{\theta \left( -p\cdot u\right) }{\exp\left({-\frac{p\cdot u - \mu }{T}}\right)+1}  \ .
\end{equation}
In the rest frame of the medium $u_\mu =(1,0,0,0)$, and thus 
\begin{equation}
\eta \left( p\cdot u\right) =\frac{\theta \left( p_0\right) }{\exp\left({\frac{p_0-\mu }{T}}\right)+1}+
\frac{\theta \left( -p_0\right) }{\exp\left({-\frac{p_0-\mu}{T}}\right)+1}  = \theta \left( p_0\right) f_f (p_0) +
\theta \left( -p_0\right)  f_{\bar{f}} (-p_0) \ , 
\end{equation}
where $f_{f, \bar{f}}$ represent the particle and antiparticle momentum distributions
$$
f_{f, \bar{f}}(E)=\frac{1}{e^{\beta\left(E \mp \mu\right)}+1}
$$
with number density
$$
n_{f, \bar{f}}=g_{f} \int \frac{d^{3} \textbf{p}}{(2 \pi)^{3}} f_{f, \bar{f}}\ .  
$$
The thermal average of a quantity $\mathcal{E}^{n}$ is 
$$
\left\langle
 \mathcal{E}^{n}_{f, \bar{f}}\right\rangle \equiv \frac{g_{f}}{n_{f, \bar{f}}} \int \frac{d^{3} \textbf{p}}{(2 \pi)^{3}} \mathcal{E}^{n} f_{f, \bar{f}}\ .
$$
Restructuring the Dirac-$\delta$ function via
\begin{align*}
\delta\left(p^{2}-m_{f}^{2}\right) &=\delta\left(\left(p-m_{f}\right)\left(p+m_{f}\right)\right) =\delta\left(p_{0}^{2}-\mathbf{p}^{2}-m_{f}^{2}\right) =\delta\left(p_{0}^{2}-\omega_{p}^{2}\right) =\delta\left(\left(p_{0}-\omega_{p}\right)\left(p_{0}+\omega_{p}\right)\right) \\
&=\frac{1}{2 \omega_{p}}\left(\delta\left(p_{0}-\omega_{p}\right)+\delta\left(p_{0}+\omega_{p}\right)\right)\ , \text{ where }\omega_p = \sqrt{\mathbf{p}^2+m^{2}_f}= E_p\ , 
\end{align*}
$ J_{n}^{(f)}$ can be written as
\begin{equation}
\begin{aligned}
J_{n}^{(f)}&=\int \frac{d^{4} p}{(2 \pi)^{3}} \delta\left(p^{2}-m_{f}^{2}\right) \eta(p \cdot u)(p \cdot u)^{n}\\
&=\int \frac{d^{3} \textbf{p}  d p_{0}}{(2 \pi)^{3}} \frac{1}{2 \omega_{p}}\left(\delta\left(p_{0}-\omega_{p}\right)+\delta\left(p_{0}+\omega_{p}\right)\right) \left(\theta \left( p_0\right) f_f (p_0) +\theta \left( -p_0\right)  f_{\bar{f}} (-p_0)\right) (p_0)^n\\
&=\int \frac{d^{3} \textbf{p}}{(2 \pi)^{3}}\left(\frac{\omega_{p}^{n}}{2 \omega_{p}} f_{f}\left(\omega_{p}\right)+\frac{\left(-\omega_{p}\right)^{n}}{2 \omega_{p}} f_{\bar{f}}\left(\omega_{p}\right)\right) \\
&=\frac{1}{2} \int \frac{d^{3} \textbf{p}}{(2 \pi)^{3}}\left(E_{f}^{n-1} f_{f}(E_p)+(-1)^{n} E_{f}^{n-1} f_{\bar{f}}(E_p)\right)\\
&=\frac{1}{2}\left(\frac{n_{f}}{g_{f}}\left\langle E_{f}^{n-1}\right\rangle+(-1)^{n} \frac{n_{\bar{f}}}{g_{\bar{f}}}\left\langle E_{\bar{f}}^{n-1}\right\rangle\right).
\end{aligned}
\end{equation}
For neutrinos with $g_\nu =g_{\bar{\nu}}=1$ we have our final results 
\begin{equation}
J_{1}^{(\nu)}=\frac{1}{2}\left(\frac{n_{\nu}}{g_{\nu}}-\frac{n_{\bar{\nu}}}{g_{\bar{\nu}}}\right)=\frac{1}{2}\left(n_{\nu}-n_{\bar{\nu}}\right) ,
\end{equation}
\begin{equation}
J^{(\nu)}_2=\frac{1}{2}\left(n_{\nu}\left\langle E_{\nu}\right\rangle+n_{\bar{\nu}}\left\langle E_{\bar{\nu}}\right\rangle\right) =\frac{7 \pi ^2 T^4}{240}\ .
\end{equation}

\end{document}